\documentclass[10pt]{emulateapj}
\usepackage{apjfonts}
\usepackage{graphicx}
\usepackage{psfig}
\slugcomment{{\sc Accepted to ApJ:} April 11, 2008}
\newcommand{\begit}{\begin{itemize}}
\newcommand{\enit}{\end{itemize}}
\newcommand{\begen}{\begin{enumerate}}
\newcommand{\enen}{\end{enumerate}}
\newcommand{\p}{\partial}    
 
\newcommand{\beq}{\begin{equation}}
\newcommand{\eeq}{\end{equation}}
\newcommand{\beqa}{\begin{eqnarray}} 
\newcommand{\eeqa}{\end{eqnarray}} 
\newcommand{\dvx}{\delta v_x}
\newcommand{\dvy}{\delta v_y}
\newcommand{\dvz}{\delta v_z}

\newcommand{\bnabla}{\mbox{$\nabla$}}

\begin{document}

\title{Gravitational Instability in Radiation Pressure Dominated Backgrounds}

\author{Todd A.~Thompson} 

\affil{Department of Astronomy and 
Center for Cosmology \& Astro-Particle Physics \\ 
The Ohio State University, Columbus, Ohio 43210 \\
thompson@astronomy.ohio-state.edu}

\begin{abstract}

I consider the physics of gravitational instabilities in the presence of dynamically important radiation pressure and gray radiative diffusion, governed by a constant opacity $\kappa$. For any non-zero radiation diffusion rate on an optically-thick scale $k^{-1}$, the medium is unstable unless the classical gas-only isothermal Jeans criterion is satisfied.  If the radiation acoustic sound crossing timescale on a scale $k^{-1}$ ($t_r$) is less than the geometric mean of the dynamical and radiative diffusion timescales ($t_{\rm diff}$ and $t_{\rm dyn}$), then diffusion is ``slow.'' In this limit, although the dynamical Jeans instability is stabilized by radiation pressure on scales smaller than the adiabatic Jeans length, on these same spatial scales the medium is unstable to a diffusive mode.  In this regime, neglecting gas pressure, the characteristic timescale for growth of this mode is independent of spatial scale and given by $(3\kappa c_s^2)/(4\pi G c)$, where $c_s$ is the adiabatic sound speed.  This characteristic timescale is that required for a fluid parcel to radiate away its thermal energy content at the Eddington limit, the Kelvin-Helmholz timescale for a radiation pressure-supported self-gravitating object. In the limit of ``rapid'' diffusion --- defined by the inequality $t_r>(t_{\rm diff} \,t_{\rm dyn})^{1/2}$ --- radiation does nothing to suppress the Jeans instability and the medium is dynamically unstable unless the gas-only Jeans criterion is satisfied. I connect with treatments of Silk damping in the early universe.  I briefly discuss several astrophysical applications, including photons diffusing in regions of extreme star formation (starburst galaxies \& pc-scale AGN disks), and the diffusion of cosmic rays in normal galaxies and galaxy clusters.  The former (particularly, starbursts) are ``rapidly'' diffusing and thus cannot be supported against dynamical instability in the linear regime by radiation pressure alone.  The latter are more nearly ``slowly'' diffusing. I speculate that the turbulence in starbursts may be driven by the dynamical coupling between the radiation field and the self-gravitating gas, perhaps mediated by magnetic fields, and that this diffusive instability operates in individual massive stars. Appendices contain a more detailed treatment of radiation transport and consideration of uniform rotation in the background medium.

\end{abstract}


\section{Introduction}
\label{section:introduction}

Standard treatments of the Jeans instability assume the medium is
homogeneous and isotropic and governed by a barotropic equation of state.
Employing the ``Jeans Swindle''
so that the Poisson equation is satisfied in an ad hoc way
with no background gradients in density, the dispersion relation
\beq
\omega^2 = c_g^2k^2 - 4\pi G\rho
\label{classical_jeans}
\eeq
follows from a linear analysis.  Here, $c_g$ is the gas
sound speed and $\rho$ is the mass density.  The Jeans instability is 
long-wavelength; for scales larger than the Jeans length, 
\beq
2\pi k_{\rm J}^{-1}=\lambda_{\rm J}=c_g(\pi/G\rho)^{1/2},
\label{classical_length}
\eeq
the system is dynamically unstable under the action of a perturbation to 
the density and the attending increase in the gravitational potential
(Jeans 1902, 1928; e.g., Binney \& Tremaine 1987). 
On scales smaller than $\lambda_{\rm J}$ the medium responds
to a compression with a restoring pressure force.
Equation 
(\ref{classical_length}) can be obtained by equating
the acoustic sound crossing timescale on a scale $\lambda_{\rm J}$ with
the dynamical timescale. 
Equivalently, equation (\ref{classical_length}) 
may be read as expressing the fact that for stability, the total 
thermal energy of the medium within a volume $\lambda_{\rm J}^3$ 
must exceed the gravitational potential energy.
When the system 
is unstable, it continuously and spontaneously transitions
to states of lower total energy by liberating thermal energy 
(e.g., Chandrasekhar 1961).


The purpose of this paper is to understand how the classical gas 
Jeans criterion is modified by radiation and to ask in which 
astrophysical environments such a modification might be important.  
Although there are many
treatments in the literature of both the Jeans instability 
(e.g., Jeans 1928; Ledoux 1951; Chandrasekhar 1954, 1958; Mestel 1965;
Lynden-Bell 1966) and the physics of radiating flows (e.g., 
Mihalas \& Mihalas 1984 and references therein; 
Spiegel 1957; Kaneko et al.~1976; Bisnovatyi-Kogan \& Blinnikov 1978, 1979; 
Mihalas \& Mihalas 1983;
Dzhalilov et al.~1992; Zhugzhda et al.~1993; 
Arons 1992; Bogdan et al.~1996; Gammie 1998; Kaneko et al.~2000;
Blaes \& Socrates 2001, 2003; Socrates et al.~2005), 
there has been relatively little work on self-gravitating
environments where radiation might play an important dynamical role
(however, see Kaneko \& Morita 2006, Vranjes \& Cadez 1990; Vranjes 1990).

Perhaps the first and most familiar treatment of self-gravitating radiation
pressure dominated media was carried out by Silk (1967), (1968),
and then extended by Peebles \& Yu (1970) and Weinberg (1971), 
in the context of acoustic wave damping of primeval fluctuations by radiative 
diffusion --- ``Silk Damping'' (see also Hu \& Sugiyama 1996; Dodelson 2003).  
However, these works focus specifically 
on the damping rate of acoustic fluctuations and the generation of entropy
and did not delineate how the Jeans criterion 
is modified on scales larger than the gas-only Jeans length when radiation
is dynamically dominant and diffusing.    They also do not discuss
the physics of slow non-dynamical diffusive modes.
%

This paper is motivated by astrophysical systems where self-gravity and 
radiation are essential. These include sites of extreme massive star formation
such as compact starburst galaxies 
and the parsec-scale disks or obscuring ``torii'' 
thought to attend the process of fueling active galactic nuclei.
These environments are marked by high radiation energy density and 
high gas density, as well as optical depths to their own dust-reprocessed 
infrared radiation that may significantly exceed unity 
(\S\ref{section:discussion} and, e.g., Pier \& Krolik 1992; 
Goodman 2003; Sirko \& Goodman 2003; Thompson et al.~2005 [TQM]; Chang et al.~2006).  
In each of these systems radiation pressure can be comparable to gravity and the 
associated photon energy density is rivaled only by the energy density in 
turbulence and, potentially, the contributions from cosmic rays and magnetic 
fields.\footnote{For a recent assessment of the strength of magnetic fields in 
starburst galaxies, see Thompson et al.~(2006). For a discussion of cosmic
ray feedback in galaxies, see Socrates et al.~(2006).} The very high radiation 
energy densities in these systems led TQM to propose a theory of marginally 
Toomre-stable radiation pressure supported starburst and AGN disks (see also
Scoville et al.~2001 \& Scoville 2003).  Additionally, this analysis may
be of some interest for the stability of individual massive stars and
for self-gravitating media whose pressure is dominated by cosmic rays,
either in normal star-forming galaxies or galaxy clusters.

In \S\,\ref{section:jeans}, I present a simple linear analysis 
of the gravitational instability.
Appendix \ref{appendix:rotation} discusses uniform rotation
in the background medium.  More detailed treatments of radiation transport 
are considered in Appendix \ref{appendix:transport} (see also Kaneko \& Morita 2006).
In \S\,\ref{section:discussion}, I discuss the relevance of the results derived
for a number of astrophysical environments
and \S\ref{section:conclusion} provides a summary.  

\section{Gravitational Instability with Radiation}
\label{section:jeans}

Here, I describe the simplest non-trivial treatment of the 
Jeans problem with radiation pressure and diffusion that captures the 
physics needed for a qualitative understanding (see Appendix \ref{appendix:transport}
for a more detailed treatment).
The equations express continuity, momentum and total energy conservation, 
self-gravity, and equilibrium optically-thick radiative diffusion 
(e.g., Mihalas \& Mihalas 1984).  They are
\beq
\frac{\p\rho}{\p t}+{\bf \bnabla\cdot}(\rho{\bf v})=0,
\label{continuity}
\eeq
\beq
\frac{\p{\bf v}}{\p t}+{\bf v\cdot \bnabla v}=-\frac{1}{\rho}{\bf \bnabla}P-{\bf \bnabla}\Phi,
\label{momentum}
\eeq
\beq
\frac{\p U}{\p t}+{\bf v\cdot \bnabla}U+(U+P){\bf \bnabla\cdot  v}=-{\bf \bnabla\cdot F},
\label{energy}
\eeq
\beq
\bnabla^2\Phi = 4\pi G\rho,
\label{gravity}
\eeq
and
\beq
{\bf F}=-\frac{c}{3\kappa\rho}{\bf \bnabla}u_r.
\label{diffusion}
\eeq
Here, $U=u_g+u_r$ and $P=p_g+p_r$ are the total internal energy density and pressure, 
and the subscripts $r$ and $g$ refer to the radiation and the gas, respectively.
The radiation pressure force $\kappa {\bf F}/c$ is contained in the $\bnabla P$ term in 
equation (\ref{momentum}).
In addition, $p_g =\rho k_B T/m_p$, $u_r=aT^4=3p_r$, $u_g = p_g/(\gamma-1)$,
$\rho$ is the gas mass density, $\gamma$ is the adiabatic index of the gas, 
${\bf F}$ is the radiative flux, and $\kappa$ is the opacity. 

For simplicity, I take $\kappa$ constant and I do not distinguish between
the Planck-, flux-, and Rosseland-mean opacities.  
The above equations also neglect the time-dependence of the radiation 
field and they assume that the radiation and gas temperatures are exactly 
equal (see Appendix \ref{appendix:transport};  see also
Mihalas \& Mihalas 1984; Gammie 1998; Blaes \& Socrates 2003; Kaneko \& Morita 2006). 
Because the Eddington approximation has been made, 
the effects of photon viscosity have been neglected (e.g., Weinberg 1971; Agol \& Krolik 1998).
In addition, in considering extreme star formation environments
where radiation is reprocessed by dust, the 
above equations neglect the two-fluid nature of the coupled dust-gas system;
that is, they assume perfect collisional and energetic coupling between the 
dust and gas (see \S\,\ref{section:discussion}).  
Finally, no terms representing
sources of optically-thin radiative heating or cooling are included.

I consider perturbations 
of the form $q \rightarrow q+\delta q \exp[i{\bf k\cdot x}-i\omega t]$,
keep only linear terms, employ the Jeans Swindle, and take the medium
and radiation field as homogeneous, isotropic, and in radiative equilibrium: 
$\rho={\rm const}$, $P={\rm const}$, $U={\rm const}$, ${\bf v}={\bf \bnabla\cdot F}={\bf F}=0$. 
The perturbed equations are 
\beq
-i\omega\delta\rho+i\rho{\bf k\cdot}\delta{\bf v}=0,
\label{wkb_continuity}
\eeq
\beq
-i\omega\delta{\bf v}+i{\bf k}(\delta P/\rho)+i{\bf k}\delta\Phi=0,
\label{wkbm}
\eeq
\beq
-i\omega\delta U+(U+P)i{\bf k\cdot}\delta{\bf v}+i{\bf k\cdot}\delta{\bf F}=0,
\label{wkbe}
\eeq
\beq
-k^2\delta\Phi-4\pi G\delta\rho=0,
\eeq
and
\beq
i{\bf k}\delta u_r+(3\kappa\rho/c)\delta{\bf F}=0.
\eeq
The thermodynamic perturbations to the total pressure and energy density are
\beq
\delta P = \left.\frac{\p P}{\p \rho}\right|_T \delta\rho+
\left.\frac{\p P}{\p T}\right|_\rho \delta T
\label{dptherm}
\eeq
\beq
\delta U = \left.\frac{\p U}{\p \rho}\right|_T \delta\rho+
\left.\frac{\p U}{\p T}\right|_\rho\delta T
\label{dutherm}
\eeq
Note that $(\p p_r/\p\rho)|_T=(\p u_r/\p\rho)|_T=0$ so that 
$(\p P/\p\rho)|_T=(\p p_g/\p\rho)|_T=c_T^2$ --- that is,
only the gas makes a contribution to the total isothermal sound speed, 
$(\p P/\p\rho|_T)^{1/2}=c_T$, in equation (\ref{dptherm}). 
The perturbation to the radiation energy density is written as
\beq
\delta u_r =4aT^3\,\delta T=A\delta T,
\eeq
where the last equality defines $A=\p u_r/\p T$.  Combining these thermodynamic
relations with the perturbation equations, one finds that
\beq
0=\left(\omega^2-c_T^2 k^2+4\pi G\rho\right) \left(1+ \frac{i\tilde{\omega}}{\omega}\right)
+(c_T^2-c_s^2)k^2,
\label{dispersion1}
\eeq
where 
\beq
\tilde\omega=\frac{ck^2}{3\kappa\rho}\left(\frac{A}{C_V}\right)
\label{rad_diff}
\eeq
is the diffusion rate on a scale $k^{-1}$, $C_V=(\p U/\p T)|_\rho$ is
the total specific heat,  $c_s^2=(\p P/\p \rho)|_s$ is the square of the 
adiabatic sound speed for the gas and radiation, $S$ is the total entropy,
and the identity 
\beq
\left.\frac{\p P}{\p \rho}\right|_S= \left.\frac{\p P}{\p \rho}\right|_T +
\left.\frac{\p P}{\p U}\right|_\rho \left[\left(\frac{U+P}{\rho}\right)-
\left.\frac{\p U}{\p \rho}\right|_T\right],
\label{relation_adiabatic}
\eeq
has been employed.
Expanding the dispersion relation and combining terms, equation (\ref{dispersion1}) becomes
\beq
\omega^3 +i\omega^2\tilde\omega
-\omega\left[c_s^2k^2-4\pi G \rho\right]
-i\tilde\omega\left[c_T^2 k^2-4\pi G\rho\right]=0.
\label{simpledispersion}
\eeq
Note that the dimensionless ratio 
\beq
\frac{A}{C_V}=\frac{\p u_r}{\p T}\left(\left.\frac{\p U}{\p T}\right|_\rho\right)^{-1}
=\left(1+\frac{u_g}{4u_r}\right)^{-1}
\label{acv}
\eeq 
that appears in equation (\ref{rad_diff}) for $\tilde\omega$ 
approaches unity in the limit $u_g/(4u_r)\rightarrow0$, and 
zero in the limit $u_g/(4u_r)\rightarrow\infty$.
Therefore, as $u_r/u_g\rightarrow0$, 
$c_s^2\rightarrow c^2_{s,\,g}=\gamma c_T^2$ and $(A/C_V)\rightarrow0$, 
equation (\ref{simpledispersion}) reduces to
\beq
\omega^3-\omega[\gamma c_T^2k^2-4\pi G\rho]=0
\label{classical_s}
\eeq
in the gas-pressure-dominated limit, fully
analogous to the classical Jeans criterion in equation (\ref{classical_jeans}),
but includes the entropy mode $\omega=0$ (e.g., Lithwick \& Goldreich 2001)
and explicitly contains
the adiabatic gas sound speed $c_{s,\,g}^2=\gamma c_T^2$. In the opposite, 
radiation-pressure-dominated limit, $u_r/u_g\rightarrow\infty$, 
$c_s^2\rightarrow c_{s,\,r}^2=(4p_r/3\rho)\gg c_T^2$.  
Neglecting gravity, the dispersion relation for radiation pressure 
acoustic waves under the assumption of optically-thick 
equilibrium radiative diffusion,  
$\omega(\omega^2+i\tilde\omega\omega-c_{s,\,r}^2k^2)\approx0$, is obtained from
equation (\ref{simpledispersion}).

\subsection{Dimensionless Numbers}

Three dimensionless numbers determine the character of the 
modes admitted by equation (\ref{simpledispersion}).  The first measures the
importance of gas pressure alone in supporting the medium on a scale $k^{-1}$:
\beq
\vartheta_T = c_T^2k^2/(4\pi G\rho).
\label{beta}
\eeq
The criterion $\vartheta_T>1$ is the classical gas-only
Jeans criterion for gravitational stability (cf.~eq.~[\ref{classical_length}]); 
$\vartheta_T$ is the ``isothermal Jeans number.'' 
The isothermal Jeans length follows by taking $\vartheta_T=1$:  
\beq
\lambda_{{\rm J},\,T}=2\pi/k_{{\rm J},\,T}=2\pi c_T/(4\pi G\rho)^{1/2}.
\label{lambdat}
\eeq
The second and third dimensionless ratios combine to determine the importance of 
radiation pressure. The first is
\beq
\vartheta_s= c_s^2k^2/(4\pi G\rho),
\label{delta}
\eeq
the ``adiabatic Jeans number,'' analogous to $\vartheta_T$,
but which includes the contribution from radiation pressure.
Taking  $\vartheta_s=1$ defines the adiabatic Jeans length:
\beq
\lambda_{{\rm J},\,s}=2\pi/k_{{\rm J},\,s}=2\pi c_s/(4\pi G\rho)^{1/2}.
\label{lambdas}
\eeq
The second ratio is
\beq
\chi=\frac{ck^2}{3\kappa\rho}\left(\frac{A}{C_V}\right)\frac{1}{(4\pi G\rho)^{1/2}},
\label{gamma}
\eeq
a measure of the diffusion rate. The limits of rapid ($\chi\gg1$)
and slow ($\chi\ll1$) diffusion are considered in Sections  
\ref{section:rapid_diffusion} and \ref{section:slow_diffusion}, respectively.

In analogy with the classical Jeans criterion, 
one might guess that if $\vartheta_s$ is larger than unity, then 
in the limit of slow diffusion the medium is stable.  This 
turns out to be false, as I show in \S\ref{section:slow_diffusion}.
In fact, if $\vartheta_T<1$ on a scale $k^{-1}$, then the medium is unstable 
regardless of $\vartheta_s$.  

Using $\xi^2=\omega^2/(4\pi G\rho)$, and the definitions for $\vartheta_T$, 
$\vartheta_s$, and $\chi$, equation (\ref{simpledispersion}) can be written as 
\beq
\xi^3+i\chi\xi^2-\xi(\vartheta_s-1)-i\chi(\vartheta_T-1)=0.
\label{dimen_disp}
\eeq

\subsection{Rapid  Diffusion}
\label{section:rapid_diffusion}

In the limit of rapid diffusion ($\chi\gg\vartheta_s,\vartheta_T,1$), the three roots 
of equation (\ref{dimen_disp}) are
\beq
\xi\approx \pm(\vartheta_T-1)^{1/2}-i\frac{(\vartheta_s-\vartheta_T)}{2\chi}
\label{high_gamma_root}
\eeq
and
\beq
\xi\approx 
-i\chi +i\frac{(\vartheta_s-\vartheta_T)}{\chi}
\label{high_gamma_diffroot}
\eeq
to first order in $\chi^{-1}$.
When $\vartheta_T>1$, the roots in equation (\ref{high_gamma_root})
correspond to stable radiation- and gravity-modified gas acoustic waves.
For large  $\vartheta_T$, these modes propagate at the isothermal sound 
speed of the gas; large $\chi$ ensures isothermality.
In the limit $\vartheta_T\rightarrow0$ and $\chi\gg\vartheta_s$,
equation (\ref{high_gamma_root}) is simply $\xi\approx\pm i$ and
the medium is dynamically unstable.  This is the classical gas-only
isothermal Jeans instability.
Note that the limit of rapid diffusion in equation (\ref{high_gamma_root})
is distinct from the high-$k$ limit, because at high-$k$ gravity, which dictates
stability/instability, disappears. However, to make an apposite 
comparison with the literature I take the high-$k$ limit and for
the acoustic modes I find that 
\beq
\omega\approx\pm c_T k-i\frac{2\kappa}{3c} u_r\left(1+\frac{3p_g}{4u_r}\right)^2,
\eeq
in agreement with Blaes \& Socrates (2003) (their eq.~[62]).  
Equation (\ref{high_gamma_diffroot}) corresponds to 
the purely damped radiation diffusion wave.

\subsection{Slow  Diffusion}
\label{section:slow_diffusion}

In the limit of slow diffusion ($\chi\ll\vartheta_s,\vartheta_T,1$),
\beq
\xi\approx\pm(\vartheta_s-1)^{1/2}-
\frac{i\chi}{2}\left(\frac{\vartheta_s-\vartheta_T}{\vartheta_s-1}\right)
\label{small_gamma_root}
\eeq
and
\beq
\xi\approx-i\chi\left(\frac{\vartheta_T-1}{\vartheta_s-1}\right).
\label{small_gamma_unstable}
\eeq
If $\vartheta_s>1$, equation (\ref{small_gamma_root})
corresponds to two stable damped gravity-modified radiation acoustic 
waves.\footnote{Note that the real part of $\xi$ is modified by $\chi$ at the level
$\xi\approx\pm(\vartheta_s-1)^{1/2}\mp(\chi^2/2)(\vartheta_s-\vartheta_T)(\vartheta_s-1)^{-3/2}$
in equation (\ref{small_gamma_root}) if the second-order term in $\chi$ is kept.}
For $\vartheta_s\gg1$ and $\vartheta_s\gg\vartheta_T$, the damping rate for 
these radiation acoustic waves is simply $\chi/2$. 
Conversely, when $\vartheta_s<1$ (and, thus, $\vartheta_T<1$)
--- that is, on scales larger than the adiabatic Jeans length (eq.~[\ref{lambdas}])---
the medium is dynamically unstable to the Jeans instability: $\xi\rightarrow\pm i$.

Equation (\ref{small_gamma_unstable}) is key.  It says that there
is an intermediate range in spatial scale $k^{-1}$, larger than 
isothermal Jeans length (eq.~[\ref{lambdat}]) and {\it smaller} than the adiabatic Jeans 
length (eq.~[\ref{lambdas}]),
that is always unstable.  For negligible gas pressure ($\vartheta_T\rightarrow0$), 
it is precisely when the adiabatic Jeans number $\vartheta_s$ is greater
than unity and the dynamical Jeans instability is suppressed in
equation (\ref{small_gamma_root}) that the diffusive mode in 
equation (\ref{small_gamma_unstable}) is unstable.  Even for 
arbitrarily large $\vartheta_s$ and small $\chi$, if the classical gas-only 
Jeans criterion indicates instability --- that is, if $\vartheta_T<1$ --- then
the medium is unstable.  

In a highly radiation pressure dominated medium with $\vartheta_T\ll1\ll\vartheta_s$, 
this diffusive mode grows at a rate
\beq
\omega\approx i\frac{4\pi G}{3\kappa}\frac{c}{c_{s,\,r}^2}
\label{slow_collapse}
\eeq
in the high-$k$ limit, independent of spatial scale.
This expression is easy to understand as the rate at which a 
self-gravitating fluid parcel 
radiates its total thermal energy content ($e\sim (4/3)u_r$) 
at the Eddington limit ($\dot{e}=4\pi G \rho c/\kappa$): 
it is the inverse of the Kelvin-Helmholz timescale: 
$t^{-1}_{\rm KH}\sim\dot{e}/e\sim 4\pi G c/(3\kappa c_{s,r}^2)$.

Equivalently, the only terms from the Euler and energy equations
that contribute to this branch of the dispersion relation are
the approximate equalities ${\bf k}\delta P/\rho\approx-{\bf k}\delta\Phi$
and $(U+P){\bf k\cdot}\delta{\bf v}\approx-{\bf k}\cdot\delta{\bf F}$
(cf.~eqs.~\ref{wkbm} \& \ref{wkbe}). Combining the energy
equation with the continuity equation,
$(\delta\rho/\rho)=({\bf k\cdot} \delta {\bf v})/\omega$,
and assuming that the medium is radiation pressure 
dominated,  
\beq
\omega=i\left(\frac{ck^2}{\kappa\rho}\right)
\left(\frac{\delta p_r}{\delta\rho}\right)
\left(\frac{1}{4}\frac{\rho}{p_r}\right).
\label{omega_1}
\eeq
On the other hand, the approximate equality 
${\bf k}\delta P/\rho\approx-{\bf k}\delta\Phi$
implies that 
\beq
\frac{\delta p_r}{\delta \rho}=\frac{4\pi G\rho}{k^2}.
\label{hydro}
\eeq
Combining
equations (\ref{omega_1}) and (\ref{hydro}), one finds
precisely equation (\ref{slow_collapse}).  Substituting into
the continuity equation, I find that 
\beq
{\bf k\cdot}\delta{\bf v}=i\frac{ck^2}{\kappa\rho}\frac{\delta p_r}{4p_r}
=i\left(\frac{4\pi G\,\delta\rho}{\kappa\rho c}\right)\left(\frac{\rho c^2}{4p_r}\right),
\eeq
which relates the density and velocity perturbations.

Therefore, in a radiation-pressure dominated medium with slow
diffusion (see \S\ref{section:slowfast}) the characteristic time
for collapse on scales smaller than the adiabatic Jeans length
is $t_{\rm KH}$, not the dynamical timescale.  Although the growth
timescale becomes long as $\kappa$ and $c_{s,r}$ become large,
for $\chi\ne0$ the medium is never formally stable if $\vartheta_T<1$.
Additionally, the existence of this instability does not require 
a pure radiation-only gas with adiabatic index of 4/3.  
Written another way, the growth timescale for this diffusive instability 
at large $k$ is 
\beq
t_{\rm KH}\approx t_{\rm diff}\left(t_{\rm dyn}/t_r\right)^2
\label{simple_growth}
\eeq
in an optically-thick, slowly-diffusing, radiation pressure dominated medium,
where  $t_r=(c_{s,\,r}k)^{-1}$ is  the radiation pressure acoustic
sound crossing timescale, and $t_{\rm diff}=3\kappa\rho/ck^2$
and $t_{\rm dyn}=(4\pi G\rho)^{-1/2}$ are the diffusion and 
dynamical timescales, respectively.  

\begin{figure*}
\centerline{\includegraphics[width=9cm]{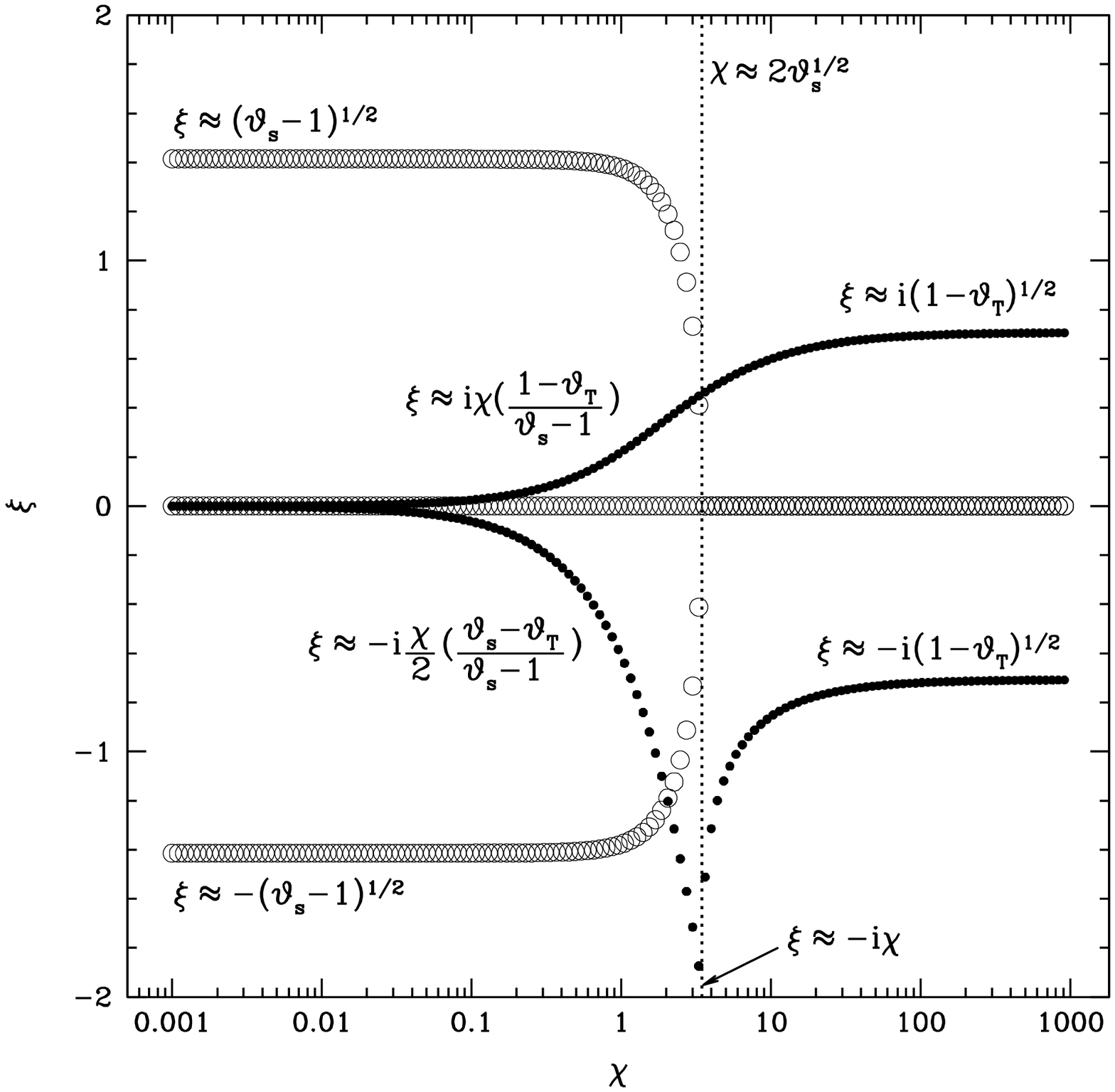} \includegraphics[width=9cm]{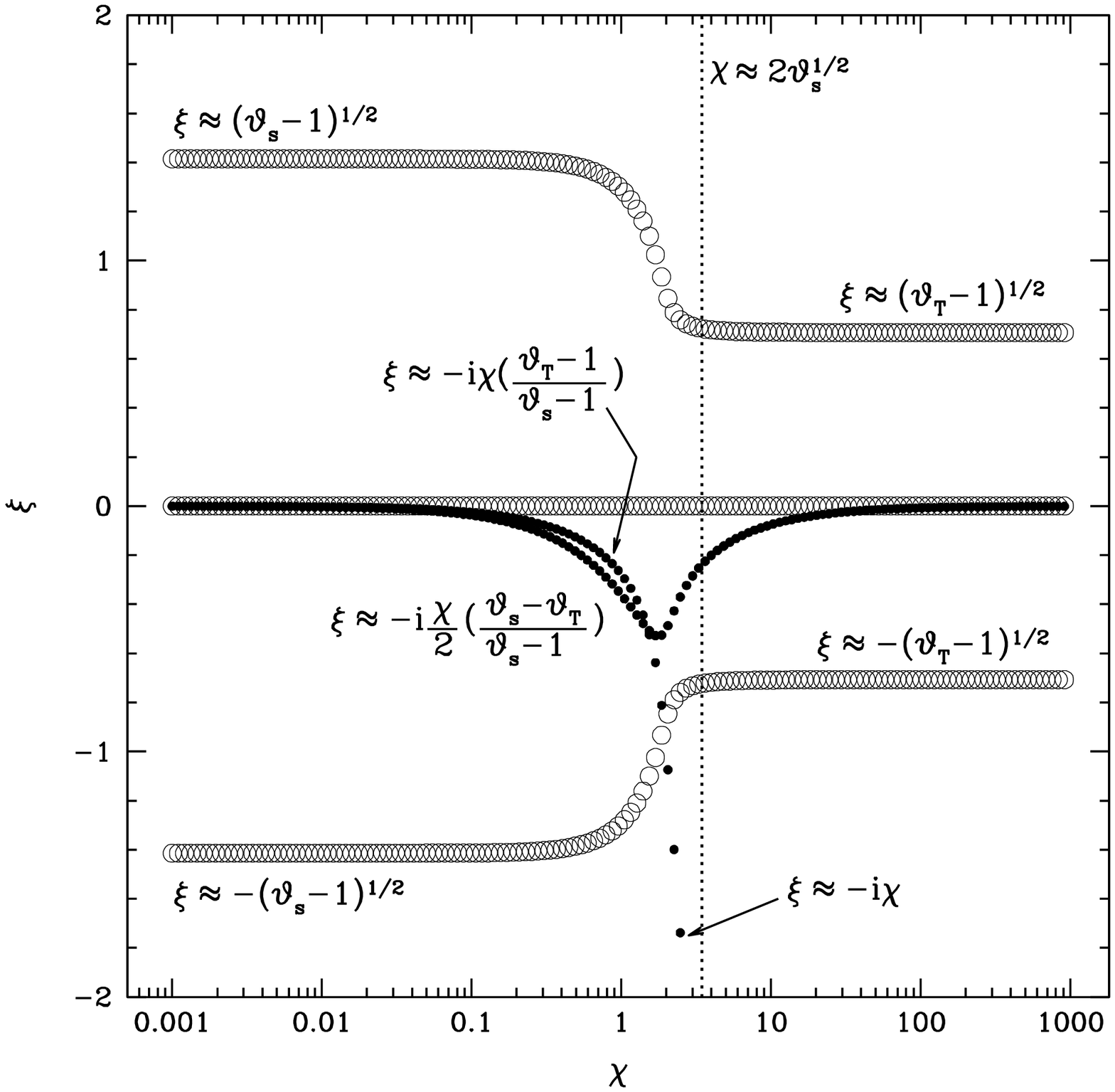}}
\figcaption[simple]{{\it Left Panel:} Solution to equation (\ref{dimen_disp})
for  $\vartheta_s=3$ and $\vartheta_T=1/2$, for $10^{-3}\le\chi\le10^3$ at fixed $k^{-1}$.
Open and filled circles show the real and imaginary parts of the three roots $\xi$, 
respectively. Positive complex roots indicate instability.  Because $\vartheta_T<1$ on the 
scale chosen, the medium is unstable any $\chi\ne0$.
For large $\chi$ the instability is dynamical,
whereas for small $\chi$ the medium is unstable to the diffusive instability
given by equation (\ref{small_gamma_unstable}).  
The dotted line shows the approximate solution for $\chi_{{\rm c},\,r}$ 
(eq.~[\ref{chi_crit}]), below which the 
gravity-modified radiation acoustic mode exists with $\xi\approx\pm(\vartheta_s-1)^{1/2}$.
For the particular parameters chosen $\chi_{{\rm c},\,r}\approx\chi_{\rm c,\,diff}$.
{\it Right Panel:} Same as the left panel, but 
for $\vartheta_T=3/2$.  Because $\vartheta_T$ is larger than unity, the gravitational instability
is stabilized for all $\chi$. Note the transition from adiabatic
(radiation plus gas) to isothermal (gas only) gravity-modified
acoustic waves. 
\label{fig:simple}}
\end{figure*}

\begin{figure*}
\centerline{\includegraphics[width=9cm]{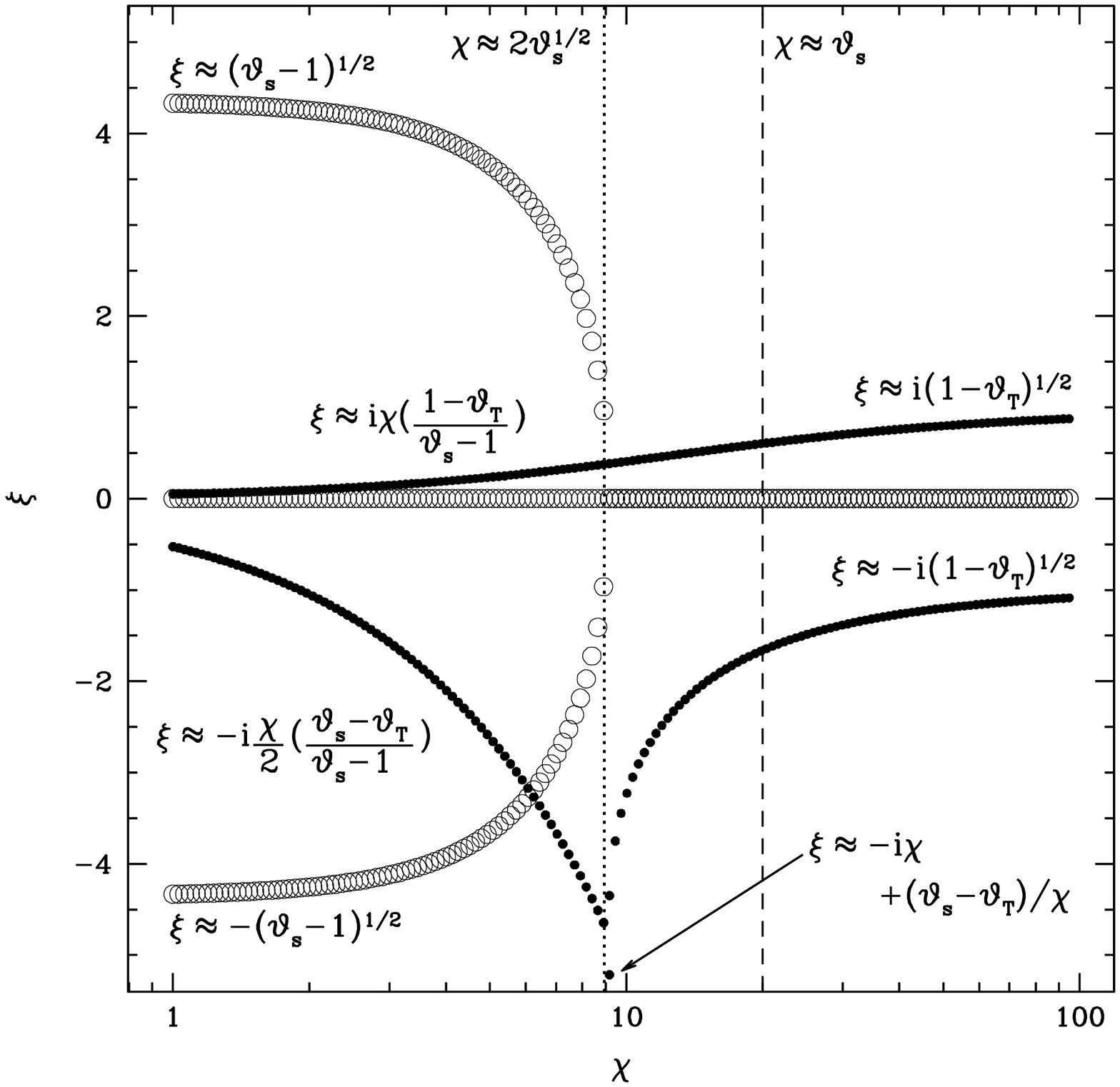} \includegraphics[width=9cm]{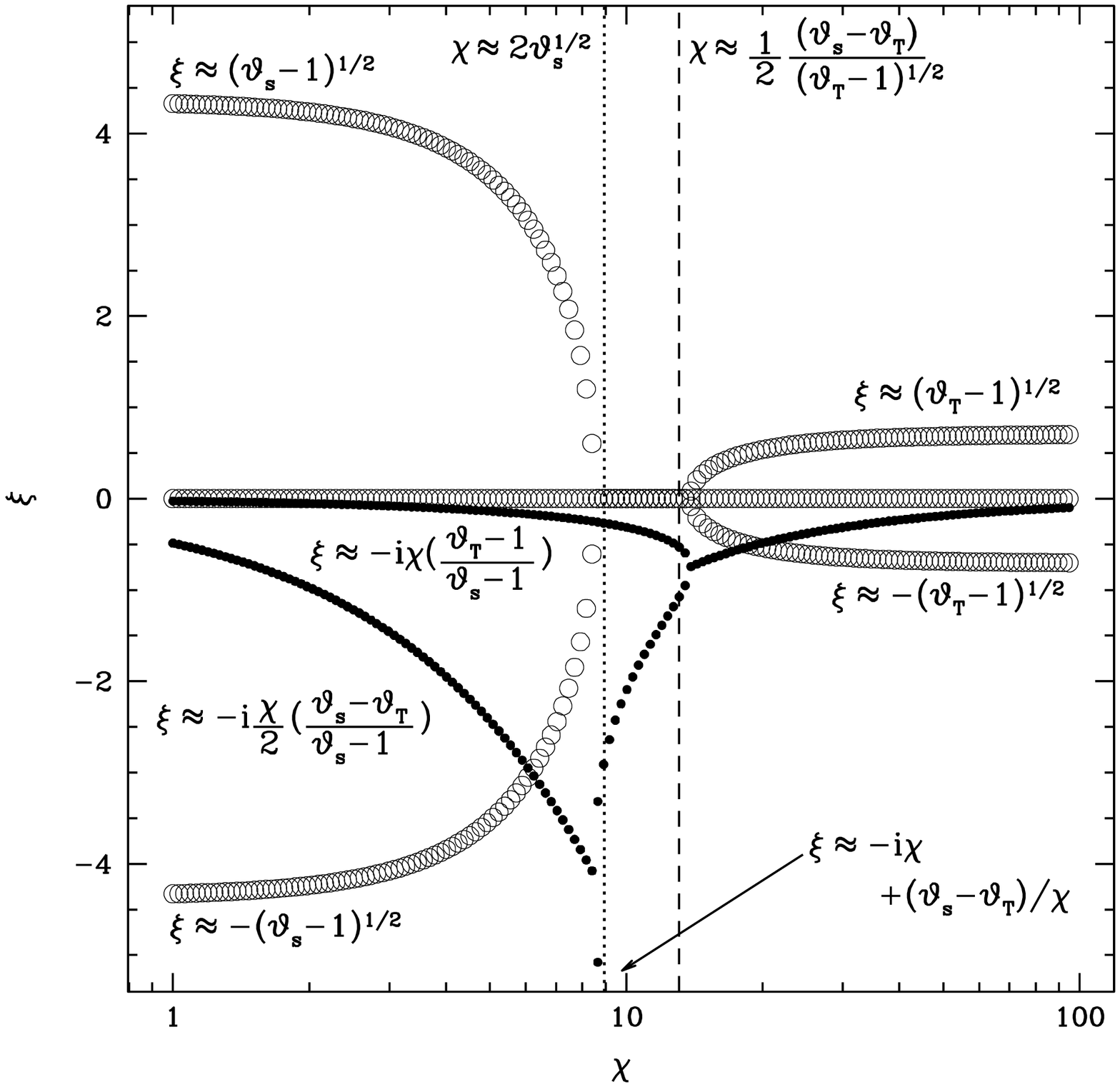}}
\figcaption[simple]{{\it Left Panel:} Same as Figure \ref{fig:simple}, but
for  $\vartheta_s=20$ and $\vartheta_T=1/10$, for $1\le\chi\le10^2$ at fixed $k^{-1}$.
Again, because $\vartheta_T<1$ on the scale chosen, the medium is unstable for any $\chi\ne0$.
For $\chi>\chi_{\rm c,\,\,diff}\approx\vartheta_s$ (dashed line) 
the instability is dynamical, whereas for small $\chi<\chi_{\rm c,\,\,diff}$ 
the growth rate for instability is $\chi(1-\vartheta_T)/(\vartheta_s-1)$ 
(eq.~[\ref{small_gamma_unstable}]).  The dotted line denotes $\chi_{{\rm c},\,r}$  
(eq.~[\ref{chi_crit}]). {\it Right Panel:} Same as the left panel, but 
$\vartheta_T=3/2$.  Because $\vartheta_T$ is larger than unity, the medium is 
stable for all $\chi$. The dashed line denotes the critical $\chi$
above which the isothermal gas acoustic wave exists with $\xi\approx\pm(\vartheta_T-1)^{1/2}$.
\label{fig:simplel}}
\end{figure*}

\subsection{Criterion for Existence of the Radiation Acoustic Mode}

If the adiabatic Jeans number is larger than unity ($\vartheta_s>1$), 
then when the radiation acoustic sound crossing timescale on a scale $k^{-1}$ 
is shorter than the diffusion timescale across that same spatial scale, the 
radiation acoustic mode can be supported by the medium.  
Thus, there is a critical diffusion rate defined by the 
rough inequality $\chi_{{\rm c},r}^{-1}\gtrsim\vartheta_s^{-1/2}$,
for which the radiation acoustic mode exists.
This criterion on the diffusion rate $\chi$ 
can be obtained from an approximate solution to equation (\ref{dimen_disp}) 
in the limit $\vartheta_s\gg1$ and $\vartheta_T\rightarrow0$,
obtained by setting $\xi\approx0$ for the radiation acoustic mode.
I find that 
\beq
1\approx\frac{\chi_{{\rm c},\,r}}{2\vartheta_s^{1/2}}
\approx\frac{2\pi}{6}\left(\frac{c}{c_{s,\,r}}\right)\frac{1}{\tau_k},
\label{chi_crit}
\eeq
where $\tau_k=\kappa\rho (2\pi k^{-1})$ is the optical depth on a spatial scale $2\pi k^{-1}$.
Thus, if $\chi$ is small with respect to $\chi_{{\rm c},\,r}$, then
the diffusion rate on a scale $k^{-1}$ is small compared to $c_{s,\,r} k$
and  the radiation acoustic mode can be supported.
Conversely, for $\chi\gtrsim\chi_{{\rm c},\,r}$, such a mode does not exist.
Note that the critical value $\chi_{{\rm c},\,r}$
is decreased by non-zero $\vartheta_T$ (see eq.~[\ref{small_gamma_root}]) and
modified if $\vartheta_s$ is larger than, but near, unity.

\subsection{Criterion Defining ``Rapid'' \& ``Slow'' Diffusion}
\label{section:slowfast}
 
Sections \ref{section:rapid_diffusion} and \ref{section:slow_diffusion}
distinguish between the limits of ``rapid'' and ``slow'' diffusion.
The criterion that separates these two limits defines a critical
diffusion rate $\chi_{\rm c,\,\,diff}$ that can be estimated by 
setting the growth timescale for the unstable diffusion mode in 
equation (\ref{small_gamma_unstable}) equal to unity, the inverse of the
dynamical timescale.  When $\vartheta_s\gg1\gg\vartheta_T$,
\beq
\chi_{\rm c,\,\,diff}\approx\vartheta_s.
\label{chi_critj}
\eeq
For $\chi>\chi_{\rm c,\,\,diff}$ the medium is ``rapidly'' diffusing 
and for $\chi<\chi_{\rm c,\,\,diff}$ it is ``slowly'' diffusing.
Alternatively, equation (\ref{chi_critj})
may be written as (cf.~eq.~[\ref{simple_growth}])
\beq
1\approx\frac{ck^2}{3\kappa\rho}\frac{(4\pi G\rho)^{1/2}}{c_{s,\,r}^2k^2}
\Longrightarrow t_r\approx(t_{\rm diff}\,\,t_{\rm dyn})^{1/2}:
\label{chi_critj2}
\eeq
if the radiation pressure acoustic
sound crossing timescale on a scale $k^{-1}$ is less than 
the geometric mean between the diffusion timescale on that same
spatial scale and the dynamical timescale, then diffusion
is ``slow'' and the stability properties of the medium 
are best described by \S\ref{section:slow_diffusion}.
Conversely, if $t_r>(t_{\rm diff} t_{\rm dyn})^{1/2}$,
diffusion is ``rapid'' (\S\ref{section:rapid_diffusion}).
This criterion is valid only at high-$k$ and in that regime is
independent of spatial scale.
Equations (\ref{chi_crit}) and  (\ref{chi_critj}) 
imply that $\chi_{\rm c,\,\,diff}$ can be greater than $\chi_{\rm c,\,\,r}$
and therefore that even though diffusion is ``slow,'' the 
radiation acoustic mode is not supported.

\subsection{Solutions to the Dispersion Relation}

The limits of fast and slow diffusion, the criterion separating them,
and the range of existence of the acoustic modes and their damping rates
are illustrated in the solution to equation (\ref{dimen_disp}) presented
in Figures \ref{fig:simple} and \ref{fig:simplel}, which show the modes $\xi$ obtained
for a wide range of $\chi$, at fixed $\vartheta_s$ and $\vartheta_T$. 
Increasing $\chi$ while keeping $\vartheta_s$ and $\vartheta_T$ constant can be 
thought of as a continuous decrease in the opacity $\kappa$ at fixed $k^{-1}$.
Open and filled circles show the real and imaginary part of $\xi$, 
respectively. Individual pieces of the various roots are labeled for comparison
with equations (\ref{high_gamma_root})$-$(\ref{small_gamma_unstable}).

The left panel of Figure \ref{fig:simple} shows a case with $\vartheta_s>1$
and $\vartheta_T<1$.  The unstable mode is 
the only positive imaginary root.  For large $\chi$ it
is the dynamical Jeans instability: $\xi\approx\pm i(1-\vartheta_T)^{1/2}$, 
whereas for small $\chi$ it is the diffusive mode of 
equation (\ref{small_gamma_unstable}).  
The dotted line shows the approximation to 
$\chi_{{\rm c},\,r}$ (eq.~[\ref{chi_crit}]).  Because for the 
parameters chosen, $\chi_{{\rm c},\,r}\approx2\vartheta_s^{1/2}\approx\vartheta_s$,
the dotted line denoting $\chi_{{\rm c},\,r}$ also roughly corresponds to 
$\chi_{\rm c,\,\,diff}$ (eq.~[\ref{chi_critj}]). 
For $\chi\lesssim\chi_{{\rm c},\,r}$, the radiation acoustic
modes are evident and modestly damped. For any $\chi\ne0$, the medium is
unstable because $\vartheta_T<1$. 
Note that the purely damped mode $\xi\approx-i\chi$ 
is off-scale for large $\chi$ (eq.~[\ref{high_gamma_diffroot}]).
Contrast the left panel of 
Figure  \ref{fig:simple} with the right panel, which shows the same calculation,
but with $\vartheta_T=3/2>1$. Because $\vartheta_T>1$, the medium
is stable for any $\chi$.  For the parameters chosen, the adiabatic
acoustic mode, which contains contributions from radiation and gas,
joins smoothly into the isothermal gas acoustic mode at modest $\chi$.

Figure \ref{fig:simplel} presents a similar calculation, but more radiation pressure dominated, 
with $\vartheta_s=20$ and $\vartheta_T=1/10$ (left panel) and $\vartheta_T=3/2$
(right panel).  The left panel of Figure \ref{fig:simple} is qualitatively
identical to the left panel of Figure \ref{fig:simplel}, but in the latter
there is a clear separation between $\chi_{{\rm c},\,r}$ (dotted line) and 
$\chi_{\rm c,\,\,diff}$ (dashed line).  The right panel of Figure \ref{fig:simplel}
again shows a case with $\vartheta_T>1$, so that the Jeans instability 
is stabilized for any $\chi$, but shows the separation in $\chi$ between
the adiabatic acoustic mode, which is here highly radiation pressure 
dominated ($\chi\lesssim\chi_{{\rm crit},\,r}$), and the damped gravity-modified
isothermal gas acoustic mode that exists for 
$\chi\gtrsim(\vartheta_s-\vartheta_T)/2(\vartheta_T-1)^{1/2}$
(dashed line).

Although I do not plot it here, equations (\ref{high_gamma_root})$-$(\ref{small_gamma_unstable})
and the general solution to equation (\ref{dimen_disp}), show that in the limit 
$\vartheta_T\rightarrow0$ and $\vartheta_s\rightarrow0$
the medium is dynamically unstable for any $\chi$.

\subsection{The Connection to Treatments of Silk Damping}

The calculation of the dispersion relation and the slow diffusion
limit of \S\ref{section:slow_diffusion} is most closely related to 
the non-relativistic calculation of Silk (1967) in the context of
the early universe.
His expression for the damping rate of adiabatic radiation-dominated
acoustic modes is essentially equivalent to the damping term in 
equation (\ref{small_gamma_root}), without the ``1'' in the denominator
and with $\vartheta_T\rightarrow0$.  The correct relativistic expression 
for the damping rate was subsequently obtained by Weinberg (1971).
The qualitative difference here with respect to Silk (1967) and
other versions of the derivation of the damping of acoustic modes
in the early universe by radiative diffusion (see, e.g., Hu \& Sugiyama 1996)
is equation (\ref{small_gamma_unstable}), which shows that the medium is 
unstable to a slow diffusive mode precisely in the regime ($\vartheta_s>1$) 
where the medium is stable to the adiabatic Jeans criterion --- that is,
on scales below the adiabatic Jeans length (eq.~[\ref{lambdas}]).  
A calculation of the dispersion
relation in the cosmological context (in an expanding background)
and an evaluation of the importance of this mode is in 
preparation.  If it grows, it should
do so very slowly on a timescale many times the dynamical timescale,
and it should be purely non-adiabatic.  Although I have not done a 
relativistic calculation here, taking $c^2_{s,\,r}=c^2/3$
in equation (\ref{slow_collapse}), I find that 
$\omega\approx i 4\pi G/(\kappa c)$.  This is the inverse of  
the Kelvin-Helmholtz timescale for a relativistic radiation
pressure supported self-gravitating object.\footnote{Note the 
correspondence with the Salpeter timescale for black hole growth.}  
The characteristic timescale for growth is of order 
$t_{\rm KH}\sim10^{16}(\kappa/0.4{\rm\,\,cm^2\,g^{-1}})$\,s, 
which is of order thousands of times longer than the
age of the universe at decoupling.  Because the total matter density 
(which sets the gravitational driving term, the numerator of 
eq.~[\ref{slow_collapse}]) is
roughly ten times the baryon density (which sets the scattering
timescale, the denominator of eq.~[\ref{slow_collapse}]), 
one expects a more careful derivation to yield a 
timescale shorter by this ratio.

\subsection{Extensions}
\label{section:extensions}

Motivated by Chandrasekhar (1961), Appendix \ref{appendix:rotation} 
contains an analysis analogous to \S\ref{section:jeans}, but including 
uniform rotation in the background medium.  As in the case with only 
gas pressure, modes propagating at right angles to the angular momentum 
vector are stabilized by rotation if the angular velocity ($\Omega$) is 
large enough that $\Omega^2>\pi G\rho$ (see also Goldreich \& Lynden-Bell 1965).

Appendix \ref{appendix:transport} 
accounts for the time-dependence of the radiation field and
the possibility of imperfect energetic coupling between the 
radiation field and the gas.   For the parameters appropriate to 
the astrophysical applications discussed in \S\ref{section:discussion}
these factors are largely unimportant for the qualitative stability 
properties of the medium.  This follows from the fact that 
the characteristic frequency for energetic coupling between
the radiation and the gas, $\omega_{\rm th}\approx\kappa\rho c (u_r/u_g)$
(see eq.~[\ref{omegath}] and surrounding discussion, as well as 
Bogdan et al.~1996; Blaes \& Socrates 2003), is likely to be large 
in the contexts considered.  As in the work of 
e.g., Dzhalilov et al.~(1992), Zhugzhda et al.~(1993), and 
Bogdan et al.~(1996) yet more precise descriptions of the 
transport should be explored, as should the dependence of
the stability properties on the temperature and density 
dependence of the opacity (e.g., Bisnovatyi-Kogan \& Blinnikov 1979;
Zhugzhda et al.~1993; Blaes \& Socrates 2003)
and the explicit dependence on the scattering albedo
(e.g., Kaneko \& Morita 2006).

Gradients in the background state --- and, particularly, a 
background flux --- have been neglected in the analysis of 
\S\ref{section:jeans}.  Ledoux (1951) considered a consistent 
background state without invoking the Jeans Swindle in calculating 
the Jeans instability and found only a small quantitative change 
to the stability properties of the medium: the Jeans length was 
increased by a factor of $\sqrt{2}$.  A detailed assessment of such gradients 
in the context of radiation pressure dominated self-gravitating media,
as well as an exploration of magnetic fields and the physics of the 
photon bubble instability (Arons 1992; Gammie 1998; 
Blaes \& Socrates 2001, 2003) are saved for a future effort.

\section{Discussion}
\label{section:discussion}

The analysis of \S\ref{section:jeans} indicates that under the assumption
of optically-thick equilibrium radiative diffusion, an isotropic self-gravitating 
medium is unstable if 
the classical gas-only isothermal Jeans criterion is not satisfied.  
If diffusion is rapid, the instability is dynamical --- the 
classical Jeans instability.  If 
diffusion is slow, then on scales larger than the adiabatic Jeans length
the medium is dynamically unstable --- again, the Jeans instability.  
However, on scales smaller than the 
adiabatic Jeans length the medium is unstable to a diffusive mode
that acts on a timescale of order the Kelvin-Helmholz 
timescale (eq.~\ref{slow_collapse}), longer than the dynamical
timescale of the medium.

Depending on whether or not the 
medium is rapidly or slowly diffusing (cf.~eqs.~\ref{chi_critj} \& \ref{chi_critj2}), 
the growth time for gravitational instability may be significantly decreased
with respect to the dynamical timescale corresponding to the 
average density of the medium.
Here I discuss several astrophysical environments where this 
analysis is applicable and where it provides some insight into the 
stability properties of the medium.

\subsection{Radiation in Starbursts, ULIRGs, 
pc-scale AGN Disks, \& Extreme Massive Star-Forming Regions}
\label{section:starburst}

Starburst galaxies are marked by high radiation energy density, 
high gas density, and optical depths to their own dust-reprocessed infrared radiation that 
exceed unity.    Scoville et al.~(2001), Scoville (2003), and TQM
have argued that radiation pressure may dominate the dynamics in these systems.
Typical temperatures are in the range $T\sim50-200$\,K
and densities range from $n\sim500-10^3$ cm$^{-3}$ (e.g., the central regions
of M82 and NGC 253) to $n\gtrsim2\times10^4$ cm$^{-3}$ (e.g., the 
nuclei of the ULIRG Arp 220; Downes \& Solomon 1998) on $\sim$100\,pc scales.
In extreme massive star-forming regions temperatures are similar
to those in starburst galaxies and ULIRGs, but the typical density of clumps
and cores responsible for star formation can be higher.
An example is the core of NGC 5253 with $n\approx10^7$ cm$^{-3}$
and a physical scale of order 1\,pc (e.g., Turner, Beck, \& Ho 2000).

Yet more extreme physical conditions are expected to obtain in the 
self-gravitating pc-scale disks or obscuring ``torii'' 
thought to attend the process of fueling active galactic nuclei
(e.g., Pier \& Krolik 1994; Goodman 2003; 
Sirko \& Goodman 2003; TQM).
There, one expects central disk temperatures approaching the sublimation
temperature of dust grains, $T_{\rm sub}\sim10^3$ K.
Although the gas density is
uncertain in these environments, we can make an estimate by assuming 
that the disk is marginally Toomre-stable such that 
$n\approx\Omega^2/(2\pi G m_p)\approx6\times10^8\,M_8R_{\rm 1\,pc}^{-3}$ cm$^{-3}$,
where $M_8=M/10^8$ M$_\odot$ and $R_{\rm 1\,pc}=R/1$\,pc, 
within the sphere of influence of the central 
supermassive black hole.\footnote{The
sublimation radius for dust is $R_{\rm sub}\approx 1L_{46}^{1/2}T_{3}^{-2}{\rm \,\,pc}$,
where $T_{3}=T_{\rm sub}/10^3$ K and $L_{46}=L_{\rm BH}/10^{46}$ ergs s$^{-1}$
is the Eddington luminosity for a $10^8$ M$_\odot$ black hole.}

Simple estimates indicate that the dust and gas in these systems
are collisionally and energetically coupled and that the IR optical
depth is larger than unity.
Ignoring the enhancement of dust-gas coupling due to magnetic fields
and grain charging, the mean free path for momentum coupling is
\beq
\lambda_{dg}\approx10^{-3} n_4^{-1}\,\,\,{\rm pc}
\approx10^{-7} n_8^{-1}\,\,\,{\rm pc},
\label{lambda_dg}
\eeq
where $n_x=n/10^x$ cm$^{-3}$.\footnote{Here, I have assumed a 
dust grain density of 3\,g\, cm$^{-3}$ and an average dust grain 
radius of $a_d\approx0.1$\,$\mu$m.
Because grain charging and magnetic fields are likely to be important 
the dust and gas may be regarded as a single, coupled fluid on scales 
larger than $\lambda_{dg}$. }
The medium is optically-thick to the dust-reprocessed IR radiation field
on scales larger than 
\beq
\lambda_{\tau=1}=(\kappa\rho)^{-1}\approx8\kappa_{2.5}^{-1}n_4^{-1}\,\,{\rm pc}
\approx8\times10^{-4}\kappa_{2.5}^{-1}n_8^{-1}\,\,{\rm pc},
\label{lambda_tau}
\eeq
where $\kappa_{2.5}=\kappa/2.5$ cm$^{2}$ g$^{-1}$ is a representative
Rosseland-mean dust opacity for $T\approx100$\,K, assuming solar metallicity and a 
Galactic dust-to-gas ratio (e.g., Figure 1 from Semenov et al.~2003).  
For temperatures near the dust sublimation temperature, $\kappa_{2.5}\approx1$
is also a fair order-of-magnitude approximation (but, see Chang et al.~2006).
Because $\lambda_{\tau=1}/\lambda_{dg}\approx8\times10^3/\kappa_{2.5}\gg1$, the dust 
and gas are always highly collisionally coupled if the average medium is optically-thick. 
A rough estimate of the optical depth in the nuclei of the ULIRG Arp 220, where the 
scale of the system is $R\sim100$ pc, is 
$\tau_{\rm IR}\approx
10\kappa_{2.5}n_4R_{\rm 100\,pc}$.
In systems like M82 and NGC 253, $n$ is lower and the medium is 
only marginally optically-thick on 100\,pc scales.  Although observations
indicate that the obscuring material surrounding AGN may occupy a large
fraction of $4\pi$, theoretical arguments suggest that most of the gas
may be confined to a thin disk with vertical scale height $h\ll R$
(e.g., Thompson et al.~2005; Chang et al.~2006; Krolik 2007).  The vertical 
optical depth in such a disk is then 
$\tau_{\rm IR}\approx10^2\kappa_{2.5}n_8 (h/{\rm 0.1\,pc})$.

Although the dust-gas fluid is highly collisionally coupled on the scales of interest, 
the energetic coupling may not be perfect.
However, in regions
for which the cooling line radiation is optically thick, we do not expect large temperature
differences between the gas and dust,  similar to the case in dense
molecular clouds where the gas temperature is maintained by a combination of 
heating by dust-gas collisions and cosmic rays, and cooling in molecular lines
(see, e.g., Gorti \& Hollenbach 2004).  At high density, inelastic dust-gas 
collisions likely dominate gas heating.
Assuming order-unity differences between the gas and dust temperatures,
the gas heating timescale is roughly
\beq
t_{\rm heat}/t_{\rm dyn}\approx0.03\,n_4^{-1/2}T_{2}^{-1/2}
\approx10^{-4}\,n_8^{-1/2}T_{3}^{-1/2},
\eeq
where here $t_{\rm dyn}=(G\rho)^{-1/2}$. 
Assuming tight dust-gas coupling
and $\tau_{\rm IR}\gtrsim1$, the medium is highly 
radiation pressure dominated:
\beq
p_r/p_g\approx10^3T_{2}^3n_4^{-1}\approx10^2T_{3}^3n_8^{-1}.
\label{pr_pg}
\eeq


Although these
estimates imply that starbursts
and AGN disks are optically-thick and potentially modestly to very strongly 
radiation pressure dominated,
the ratio\footnote{Here,  $(A/C_V)$ in $\chi$ is taken as $\approx1$ 
(see eqs.~[\ref{acv}], [\ref{gamma}]).}
\beq
\frac{\chi}{\vartheta_s}\approx\frac{(4\pi G\rho)^{1/2}}{(4p_r\kappa/c)} 
\approx1400\kappa_{2.5}^{-1}n_4^{1/2}T_{2}^{-4} 
\approx14\kappa_{2.5}^{-1}n_8^{1/2}T_{3}^{-4}
\label{dimen_starburst}
\eeq
shows explicitly that diffusion of radiation in starbursts is very ``rapid.''
The stability properties of these media on the scale of the system
are thus best represented by the far right-hand 
portion of the left panels of Figures \ref{fig:simple} and \ref{fig:simplel}: 
gas pressure is negligible and radiation pressure is 
important as measured by the adiabatic Jeans number $\vartheta_s$, 
but $\chi\gg \vartheta_s$ and the medium is dynamically (Jeans) unstable.  

For fiducial parameters, the scaling for AGN disks 
also indicates that they are rapidly diffusing.  However, the disk parameters
in this regime are quite uncertain.  For example,
Chang et al.~(2006) advocate $\kappa_{2.5}\approx20$ or larger for a gas density 
$n\sim10^8$ cm$^{-3}$ and solar metallicity, implying $\chi/\vartheta_s\sim0.7T_3^{-4}$.
The linear dependence of $\kappa$ on metallicity and the
strong temperature dependence of $\chi/\vartheta_s$ implies that if grains persist
for $T\gtrsim10^3$\,K and/or the composition of the disk is super-solar, 
the medium may transition to slowly diffusing and the timescale for 
gravitational instability 
will be increased to $t_{\rm KH}$ (eqs.~[\ref{slow_collapse}] \& [\ref{simple_growth}]).
Also, an estimate of $\vartheta_T$ shows that gas 
pressure becomes important on small scales in AGN disks and may
stabilize the medium in the linear regime (TQM).

\subsubsection{The Non-Linear Outcome}
\label{section:nonlinear}

Consider an initial hypothetical equilibrium 
configuration for a self-gravitating disk with starburst/ULIRG-like 
characteristics such that $\vartheta_T\ll1$, 
$\vartheta_s\approx1$ on $\sim100$\,pc scales and with 
$c_s^2\approx c_{s,\,r}^2\approx p_r/\rho\approx (h\Omega)^2$, where 
$h$ is the disk scale-height (as in the models of TQM).
Because diffusion is ``rapid,''   
the configuration is dynamically unstable on all scales larger
than the classical gas Jeans length ($\ll h$) and vertical hydrostatic equilibrium
cannot be maintained in the linear regime.\footnote{The equilibrium imagined
is likely also unstable to convective, magneto-rotational, and 
photon-bubble instabilities (see Blaes \& Socrates 2001, 2003).}

The non-linear outcome of the Jeans instability in such a  system is highly 
uncertain in part because it is tied to star formation, which, in turn, 
determines the character of the radiation field.  The question of whether
or not hydrostatic equilibrium can be maintained depends crucially on 
the non-linear coupling of the radiation and the gas. 
One possibility is that the large-scale radiation field
produced by star formation is coupled to the generation of turbulence,
which regulates the structure of the galaxy and its stability
properties.  
Super-sonic turbulence on large scales has been shown to 
inhibit the Jeans instability and gravitational
collapse (e.g., Klessen et al.~2000; MacLow \& Klessen 2004).  
It has also recently been invoked as a basis for understanding
the origin of the Schmidt/Kennicutt laws (Kennicutt 1998; Krumholz \& McKee 2005).
Indeed, the turbulent velocities inferred in local starbursts and ULIRGs
are large enough that a ``turbulent Jeans number,'' 
$\vartheta_{\rm turb}\approx\delta v^2 k^2/(4\pi G\rho)$,
analogous to $\vartheta_s$ and $\vartheta_T$, may indicate marginal stability: 
$\vartheta_{\rm turb}\approx1$.
Thus, if radiation pressure forces can generate turbulence, perhaps
a statistical hydrostatic equilibrium can be maintained.
 
There are at least two reasons why --- in the absence of 
energetic input to the ISM from stars (e.g., supernovae) --- that star formation
may be coupled to the generation of turbulence.  First, 
because in the initial fictitious equilibrium state envisioned radiation pressure 
is large enough that $p_r/\rho\approx (h\Omega)^2$,
the radiation field is capable of driving mass motions with velocities
of order $\delta v \approx h \Omega$ if order-unity spatial variations in the 
radiation field are present.  Second, although I have not shown it in this paper,
one expects the astrophysical environments described here to be subject
to the self-gravitating analog of the photon bubble instability, 
which in its non-linear state will
drive turbulence (Turner et al.~2005).  The latter is particularly
interesting because it motivates a dynamical coupling between 
the turbulent energy density ($u_{\rm turb}$), the photon energy density
($u_{\rm ph}$), and the magnetic energy density ($u_B$).

The second of these connections, between $u_{\rm ph}$ and $u_B$ in galaxies, 
can be motivated phenomenologically.
Recently Thompson et al.~(2006) have shown that the
magnetic field strengths in starbursts significantly exceed estimates 
derived from the ``minimum energy argument.'' In addition, they show that
$u_B$ must be a constant, order unity, multiple of $u_{\rm ph}$ in these
systems (see Condon et al.~1991).  This conclusion follows from the linearity of 
the FIR-radio correlation, the radio spectral indices of star-forming
galaxies at GHz frequencies, and the fact that the ratio
$u_B/u_{\rm ph}$ measures the importance of 
synchrotron versus inverse Compton cooling of the cosmic ray electrons
and positrons (e.g., Condon 1992). The fact the galaxies that comprise
the FIR-radio correlation have $u_{\rm ph}$'s that span five to six dex, 
and that in the Galaxy $u_{\rm ph}\approx u_B$,
implies that $u_B$ must increase with $u_{\rm ph}$, 
from normal Milky Way-like galaxies to ULIRGs.
The necessity of this lock-step increase in both $u_{\rm ph}$
and $u_B$ may signal a dynamical coupling between the radiation field 
and the magnetic field in galaxies.  
Thus, the fact that $u_{\rm ph}$, $u_B$, and $u_{\rm turb}$ are of the same order
of magnitude may not be a coincidence, but instead a necessary consequence of the 
dynamical coupling between the radiation field and the self-gravitating 
magnetized ISM.

\subsection{Cosmic Rays in Normal Star-Forming Galaxies
\& Clusters}
\label{section:cosmic_rays}

Although no attempt is made here to model the diffusion of 
cosmic rays, it is instructive to consider the 
various parameters governing gravitational stability in the case of cosmic
rays vis \`a vis radiation.

The total pressure in cosmic rays in the Galaxy is $p_{\rm cr,\,MW}\approx10^{-12}$\,ergs 
cm$^{-3}$ (e.g., Boulares \& Cox 1990), comparable to the energy density
in starlight, magnetic fields, and turbulence. The cosmic ray lifetime is 
inferred to be $t_{\rm cr}\approx2-3\times10^7$ yr (Garcia-Munoz et al.~1977; Connell 1998).
Interpreted as a diffusion timescale on kpc scales, one infers a cosmic ray 
scattering mean free path of order $l_{\rm mfp}\approx0.1-1$\,pc.  
Additionally, from the observed grammage traversed by
cosmic rays in the Galaxy, one infers an average gas density 
encountered by the cosmic rays of $n\approx 0.2$ cm$^{-3}$
(Engelmann et al.~1990; Webber et al.~2003).
Writing $l_{\rm mfp}=(\kappa\rho)^{-1}$
(cf.~eqs.~[\ref{dimen_starburst}]; see also Kuwabara \& Ko 2004),
$\vartheta_s\approx50\,\,p_{12}\,n_{0.2}^{-2}\lambda_{\rm kpc}^{-2}$
%
and\footnote{For simplicity, here I take $A/C_V=1$ in $\chi$.  This is a rough
approximation in the context of the Galaxy because, depending on which
gas phase of the ISM is being considered, $\vartheta_T$ may be the same order of 
magnitude as $\vartheta_s$.}
\beq
\frac{\chi}{\vartheta_s}\,\,\approx\frac{c \rho l_{\rm mfp}}{4p_{\rm cr}}(4\pi G\rho)^{1/2}
\,\,\approx\,\,0.4\,l_{\rm 0.1\,pc}p_{\rm cr, \,MW}^{-1}\,n_{0.2}^{3/2}.
\label{dimen_cr}
\eeq
Although the parameters are uncertain, equation (\ref{dimen_cr}) indicates
that cosmic rays are marginally slowly diffusing in normal star-forming
galaxies. Thus, as for photons in 
the dense pc-scale AGN disks discussed in \S\ref{section:starburst}, perhaps 
on kpc scales $\chi/\vartheta_s$ may be somewhat less than unity so that the growth
rate for the gravitational instability is (cf.~\ref{simple_growth}) 
$$t_{\rm KH}/t_{\rm dyn}\sim3p_{\rm cr, \,MW}\,l^{-1}_{0.1\,{\rm pc}}n_{0.2}^{-1/2}.$$
Depending on the phase of the ISM considered, the isothermal Jeans 
number $\vartheta_T$ may be very close to unity so that the critical $\chi$ below which 
the Jeans instability is suppressed can be made significantly larger 
(eq.~[\ref{small_gamma_unstable}]).  Indeed, cosmic rays have
recently been proposed as an important large-scale feedback mechanism in 
star-forming galaxies (Socrates et al.~2006).
 
Similar estimates may be written down for the central regions of
galaxy clusters, but it is unclear if these regions
may plausibly be cosmic ray pressure dominated. If they
are at least modestly so, scaling from equation (\ref{dimen_cr})
for higher pressures and lower densities,
we see then that for $0.1\lesssim l_{\rm mfp}\lesssim100$\,pc
they are plausibly in the ``slow'' diffusion limit  ($\chi/\vartheta_s<1$);
again, the dynamical Jeans instability is quelled by the non-thermal pressure 
support.  In this limit, the diffusive instability identified in equation 
(\ref{small_gamma_unstable}) still acts on $t_{\rm KH}$, but this
timescale is likely many times the age of the universe:
for $c_{s,\,r}\approx1000$\,km s$^{-1}$ and $n\approx10^{-2}$,
$t_{\rm KH}\approx10^{12}\,(l_{\rm mfp}/{\rm pc})$\,yr.

\subsection{Individual Massive Stars}

Individual massive stars are radiation pressure dominated and 
slowly diffusing and may in principle also be subject to 
the diffusive mode identified in \S\ref{section:jeans}. 
This is simply a secular instability of the kind discussed in,
e.g., Hansen (1978), and references therein.
If present, the growth timescale is the Kelvin-Helmholz time, of 
order $t_{\rm KH}\sim 3\kappa c_{s,\,r}^2/(4\pi Gc)\sim10^3$\,yrs for typical
parameters, where $\kappa$ is the Thomson opacity and $c_{s,\,r}$ is the 
adiabatic radiation pressure dominated sound speed of the fluid $\sim (GM/R)^{1/2}$. 
Although massive stars are known to be globally secularly unstable on the 
Kelvin-Helmholtz timescale, it is possible that otherwise 
stably-stratified (radiative) regions of their interiors may be 
locally unstable to a variant of the diffusive instability
in equation (\ref{slow_collapse}).


\section{Summary \& Conclusion}
\label{section:conclusion}

I consider the physics of gravitational instabilities in
the presence of dynamically important radiation pressure and 
radiative diffusion.  I find that the medium is always stable on scales
smaller than the gas-only isothermal Jeans length, $\lambda_{{\rm J}, \,T}$
(eq.~[\ref{lambdat}]). For scales larger than $\lambda_{{\rm J}, \,T}$ there
are two possibilities depending on whether the medium is ``slowly'' or
``rapidly'' diffusing, as defined in \S\ref{section:slowfast}.
When diffusion is rapid,  radiation leaks out of a perturbation 
without providing a sufficient restoring pressure force and the medium is 
dynamically unstable on all scales larger than $\lambda_{{\rm J}, \,T}$,  
regardless of the dominance of radiation pressure.

The limit of slow diffusion is more interesting.  Here, the medium 
is unstable to a diffusive mode at an intermediate range of scales
between the gas-only isothermal Jeans length $\lambda_{{\rm J}, \,T}$
and the larger (gas $+$ radiation) adiabatic Jeans length $\lambda_{{\rm J}, \,s}$ 
(eq.~[\ref{lambdas}]).  The characteristic growth timescale is 
longer than the dynamical timescale.  Neglecting gas pressure,
it is given approximately by equation (\ref{slow_collapse}) 
(see also eqs.~[\ref{small_gamma_unstable}] \&[\ref{simple_growth}]),
which is simply the Kelvin-Helmholz timescale for a radiation pressure
supported self-gravitating fluid parcel to radiate its total thermal 
energy at the Eddington limit.  Note that on small spatial scales, the
characteristic timescale is independent of scale.
For $\lambda>\lambda_{{\rm J}, \,s}$ 
the medium is dynamically ``Jeans'' unstable, as expected.  Thus, even when 
radiation pressure is dynamically dominant, on precisely the scales 
where the medium is dynamically stable by the usual Jeans criterion
($\lambda<\lambda_{{\rm J}, \,s}$) 
it is unstable to a diffusive instability that operates on the 
Kelvin-Helmholtz time.  I conclude that radiation cannot formally
stabilize a self-gravitating medium on scales larger than the 
gas-only isothermal Jeans length.  See also the discussion of
Kaneko \& Morita (2006).


In \S\ref{section:starburst}, I consider the importance of the 
results derived in \S\ref{section:jeans} for extreme sites of massive
star formation including starburst galaxies and pc-scale AGN disks.  
I argue that the average medium in these 
systems is likely to be radiation pressure dominated and 
optically-thick.  Importantly, for fiducial parameters 
the photons in these systems are in the rapidly diffusing limit 
($\chi/\vartheta_s>1$; eq.~[\ref{dimen_starburst}]).  
For fairly extreme choices for the uncertain physical 
parameters in pc-scale AGN disks (e.g., the opacity $\kappa$)
this environment is marginally slowly diffusing and thus the 
stability properties of the medium might be qualitatively different
from rapidly-diffusing starbursts.

TQM developed a theory of marginally Toomre-stable
starburst and AGN disks supported by feedback from radiation pressure. 
Because the IR photons produced by dust-reprocessed 
starlight in these systems diffuse rapidly and because their characteristic sizes  
are much larger than the classical gas-only Jeans length, the analysis
presented here dictates that they cannot be supported 
in the linear regime by radiation pressure alone.  One may wonder, then,
why the entire mass of gas in starbursts does not fragment into stars on
a single dynamical time, in apparent contradiction with observations
(e.g., Kennicutt 1998).  If radiation pressure is to be the dominant
feedback mechanism, then the answer must be that these forces are 
coupled to the generation of supersonic turbulence, which may mitigate against 
complete collapse and fragmentation on scales larger than the gas-only
Jeans length.  In \S\ref{section:nonlinear} I argue that 
the generation of turbulence likely proceeds from the non-linear coupling
of the Jeans instability with the radiation field through star formation,
and may be driven by the self-gravitating analog of the photon bubble 
instability.   This may help explain the apparent order-of-magnitude
equivalence between the radiation, magnetic, and turbulent energy densities
in starburst systems.  Thus, it is important to emphasize 
that the conclusion that radiation pressure alone cannot stave off 
gravitational instability in the rapidly diffusing limit 
does not necessarily imply that 
the disk cannot be maintained in global hydrostatic equilibrium
in an average sense by radiation pressure in the non-linear, turbulent regime.

In \S\ref{section:cosmic_rays}, I consider the case of cosmic rays 
diffusing in the Galaxy and the cosmic ray halo, and in galaxy clusters.  
Although no attempt is made here to calculate the physics of cosmic 
ray diffusion and their thermal coupling to the gas, they
provide a useful point of contrast with radiation because of 
their very high scattering optical depths.  Even so, this 
constituent of the ISM of the Galaxy is at the border of ``slow'' 
and ``rapid'' diffusion ($\chi/\vartheta_s\approx1$) 
outlined in \S\ref{section:jeans} (eq.~[\ref{dimen_cr}]) for fiducial parameters.
In the cluster context diffusion is likely more fully in the ``slow'' limit,
but it is unclear if cosmic rays dominate the total pressure budget in the
central regions (e.g., Guo \& Oh 2008).  
Finally, I also briefly mention the possibility that individual
massive stars may be locally unstable to this diffusive mode 
on the local Kelvin-Helmholz timescale in otherwise stably-stratified 
radiative regions of their interiors.

\vspace*{.1cm}

\acknowledgments 

This paper was motivated in part by stimulating conversations
with Aristotle Socrates.  I also thank Yoram Lithwick, Andrew Youdin, 
Charles Gammie, Kristen Menou, and Jeremy Goodman
for several useful conversations and Eliot Quataert, 
Norm Murray, Bruce Draine, and Julian Krolik for encouragement.
Finally, I am grateful to the Department of Astrophysical Sciences
at Princeton University, where much of this work was completed.  
This paper is dedicated to Garnett A.~B.~Thompson.

\appendix

\section{Uniform Rotation}
\label{appendix:rotation}

Chandrasekhar (1954) and (1961) explored the effect of uniform rotation 
on the Jeans instability and found that the dispersion relation is 
modified by the Coriolis force in the rotating frame.  In particular,
he showed that for the special case of waves propagating at right angles 
to the direction $\hat\Omega$ ($\cos^2\theta=0$) that equation (\ref{classical_jeans}) 
becomes $\omega^2=4\Omega^2+c_g^2k^2-4\pi G\rho$ --- that is, if 
$\Omega^2>\pi G\rho$, then the Jeans instability is stabilized for any gas 
sound speed $c_g$ ($\delta p_g=c_g^2\delta\rho$ assumed). 
For the general case $\cos^2\theta\ne0$ --- for waves whose wave vectors have arbitrary 
angles with respect to the spin axis ---  
if the classical gas Jeans criterion (eq.~[\ref{classical_length}]) indicates 
instability ($\vartheta_T<1$), then the medium is unstable for any $|\Omega|$.  
In order to gain some intuition and to make contact with
the work of Chandrasekhar (1954) it is useful to consider the
Jeans instability including uniform rotation, radiation pressure,
and radiative diffusion. The equation expressing conservation of momentum 
in the rotating frame is
\beq
\frac{\p{\bf v}}{\p t}+{\bf v\cdot \bnabla v}=
-\frac{1}{\rho}{\bf \bnabla}P-{\bf \bnabla}\Phi+
2({\bf v \times \Omega}).
\label{momentum_r}
\eeq
All other equations in the original analysis of \S\,\ref{section:jeans}
are unchanged.
I take ${\bf k}=(0,0,k_z)$ and ${\bf\Omega}=(0,\Omega_y,\Omega_z)$.  The 
perturbation equations in component form are
\beqa
-i\omega\delta\rho+i\rho k_z\dvz&=&0 \nonumber\\
-i\omega\dvx-2\dvy\Omega_z+2\dvz\Omega_y&=&0 \nonumber\\
-i\omega\dvy+2\dvx\Omega_z&=&0 \nonumber\\
-i\omega\dvz+ik_z\delta P/\rho+ik_z\delta\Phi-2\dvx\Omega_y&=&0 \nonumber\\
-i\omega\delta U+(U+P)ik_z\dvz+ik_z\delta F_z&=&0\nonumber \\
-k^2\delta\Phi-4\pi G \delta\rho&=&0\nonumber \\
ik_z\delta u_r+3\kappa\rho\delta F_z/c&=&0.\nonumber \\
\eeqa
The resulting dispersion relation is (cf.~eq.~[\ref{simpledispersion}])  
\beqa
\omega^5&+&i\tilde\omega\omega^4
-\omega^3\left[4\Omega^2+c_s^2k^2-4\pi G\rho\right]
-i\tilde\omega\omega^2\left[4\Omega^2+c_T^2k^2-4\pi G\rho\right] \nonumber \\ 
&+&\omega(4\Omega^2\cos^2\theta)\left[c_s^2k^2-4\pi G\rho\right] 
+i\tilde\omega(4\Omega^2\cos^2\theta)\left[c_T^2k^2-4\pi G\rho\right]=0,
\label{rotation}
\eeqa
where $\cos\theta=\Omega_z/|\Omega|$ and $\tilde\omega$ is the radiation
diffusion rate given in equation (\ref{rad_diff}). Defining
\beq
Q=\Omega^2/(\pi G\rho),
\eeq
and using the definitions for $\vartheta_T$, $\vartheta_s$, $\chi$, and $\xi$, in 
equations (\ref{beta})$-$(\ref{gamma}), equation (\ref{rotation}) can be rewritten as  
\beq
\xi^5\,\,+\,\,i\chi\xi^4\,\,-\,\,\xi^3\left(Q+\vartheta_s-1\right)
\,\,-\,\,i\chi\xi^2\left(Q+\vartheta_T-1\right) 
\,\,+\,\,\xi Q \cos^2\theta\left(\vartheta_s-1\right)
\,\,+\,\,i\chi Q \cos^2\theta \left(\vartheta_T-1\right)=0,
\label{rotation_disp}
\eeq
For $Q=0$, equation (\ref{rotation_disp}) reduces to equation (\ref{dimen_disp}).
In addition, for the special case $\cos^2\theta=0$ equation (\ref{rotation_disp})
reduces to equation (\ref{dimen_disp}) with the substitutions
$\vartheta_T\rightarrow Q+\vartheta_T$ and $\vartheta_s\rightarrow Q+\vartheta_s$.
Therefore, for $Q\ge1$ and $\cos^2\theta=0$, the medium is stabilized for 
any $\chi$, $\vartheta_T$, and $\vartheta_s$.  As in Chandrasekhar (1961),
I find that for all $\cos^2\theta\ne0$, if $\vartheta_T<1$, the medium is unstable.


\section{More General Treatments of Radiation Transport}
\label{appendix:transport}

The prescription for radiation transport in \S\,\ref{section:jeans} makes
several approximations.  In particular, it neglects the time-dependence 
of the radiation field and it assumes perfect radiative equilibrium
so that the radiation and gas temperatures are identical.  
The latter assumption is particularly suspect when diffusion is 
rapid on a scale $k^{-1}$, since radiative equilibrium may not be possible to maintain.
In fact, contrary to the results of \S\,\ref{section:jeans}, when the 
radiation and gas temperatures are distinguished, the gas acoustic speed in
the limit of rapid diffusion should be the adiabatic gas sound speed and not the 
isothermal gas sound speed (e.g., Mihlalas \& Mihalas 1984).  
More detailed treatments of radiating flows without self-gravity 
may be found in Dzhalilov et al.~(1992), Zhugzhda et al.~(1993), and
Bogdan et al.~(1996).  Kaneko \& Morita (2006) provide a detailed 
treatment of the radiation that distinguishes between scattering
and pure absorptive opacity.

For completeness, here I present an analysis similar to \S\ref{section:jeans},
but including the dynamics of the radiation field and allowing 
for energetic decoupling between the radiation and gas.
The set of equations is (cf.~eqs.~[\ref{continuity}]$-$[\ref{diffusion}])
\beqa
\frac{D\rho}{Dt}+\rho{\bf \bnabla\cdot}{\bf v}&=&0 \nonumber \\
\frac{D{\bf v}}{D t}+\frac{1}{\rho}{\bf \bnabla}p_g
+{\bf \bnabla}\Phi-\kappa{\bf F}/c&=&0 \nonumber \\
\frac{D u_g}{Dt}+\gamma u_g{\bf \bnabla\cdot  v}
-\kappa\rho c( u_r- a T^4)&=&0\nonumber \\
\frac{D u_r}{Dt}+\frac{4}{3}u_r{\bf \bnabla\cdot  v}
+\bnabla\cdot{\bf F}+\kappa\rho c (u_r-a T^4)&=&0\nonumber \\
\bnabla^2\Phi -4\pi G\rho&=&0\nonumber \\
\frac{1}{c}\frac{D{\bf F}}{D t}
+\frac{c}{3}\bnabla u_r+\kappa\rho {\bf F}&=&0,
\label{general_transport}
\eeqa
where $D/Dt=\p/\p t+{\bf v}\cdot\bnabla{\bf v}$ is the Lagrangian derivative, 
$T$ is the gas temperature, $a=4\sigma_{\rm SB}/c$ is the radiation energy density 
constant, and $\gamma$ is the adiabatic index of the gas.  The time-derivative of
the flux in the Euler equation has been neglected.
The perturbation equations are
\beqa
-i\omega\delta\rho+i\rho{\bf k\cdot}\delta{\bf v}&=&0\nonumber \\
-i\omega\delta{\bf v}+i{\bf k}(\delta p_g/\rho)+i{\bf k}\delta\Phi-
\kappa\delta{\bf F}/c&=&0\nonumber \\
-i\omega\delta u_g+\gamma u_gi{\bf k\cdot}\delta{\bf v}+
\kappa\rho c(\delta u_r-A_g\delta T)&=&0\nonumber \\
-i\omega\delta u_r+(4/3)u_r i{\bf k\cdot}\delta{\bf v}+i{\bf k\cdot}\delta{\bf F}-
\kappa\rho c(\delta u_r-A_g\delta T)&=&0\nonumber \\
-k^2\delta\Phi-4\pi G\delta\rho&=&0 \nonumber \\
-i\omega\delta{\bf F}+(c^2/3)i{\bf k}\delta u_r+(\kappa\rho c)\delta{\bf F}&=&0.
\label{general_perturb}
\eeqa
where I have taken $\delta(a T^4)=4aT^3\delta T=A_g\delta T$ and
employed the Jeans Swindle.
The perturbations to the gas pressure and energy density 
are written as $\delta p_g=c_T^2\delta\rho+(\p p_g/\p T)|_\rho\delta T$
and $\delta u_g=c_T^2/(\gamma-1)\delta\rho+(\p u_g/\p T)|_\rho\delta T$.
Solving equations (\ref{general_perturb}), 
the resulting dispersion relation can be written in a number of ways.  The 
form most conducive to comparison with equation (\ref{simpledispersion}) is perhaps
\beqa
\omega^5&+&i\Gamma\omega^4(2+s)
-\omega^3\left[\Gamma^2(1+s)+\nu^2+\gamma c_T^2k^2+
\frac{4}{9}\frac{u_r}{\rho}k^2-4\pi G\rho\right] \nonumber \\
&-&i\Gamma\omega^2\left[\nu^2 s+\gamma c_T^2 k^2\left(\frac{4s}{3\gamma}+2\right)
+\frac{4}{9}\frac{u_r}{\rho}k^2\left(3\gamma-2+s\right)-(2+s)4\pi G\rho\right] \nonumber \\
&+&\Gamma^2\omega\left[\gamma c_T^2k^2\left(\frac{4s}{3\gamma}+1+\frac{\nu^2}{\Gamma^2}\right)
+\frac{4}{9}\frac{u_r}{\rho}k^2\left(3\gamma-3+s\right)
-4\pi G\rho\left(1+s+\frac{\nu^2}{\Gamma^2}\right)\right]\nonumber \\
&+&i\Gamma s\nu^2\left(c_T^2 k^2-4\pi G\rho\right)=0,
\label{general_flux}
\eeqa
where 
\beq
s=\frac{A_g}{\p u_g/\p T|_\rho}=\frac{4aT^4}{u_g},
\hspace{.5cm}
\nu^2=\frac{c^2k^2}{3},
\hspace{1cm}
\Gamma=\kappa\rho c,
\hspace{.5cm}{\rm and}\,\,\,\,
\frac{\nu^2}{\Gamma^2}=\frac{1}{3}\frac{1}{(\kappa\rho k^{-1})^2}=\frac{1}{3}\frac{1}{\tau_k^2}.
\eeq
The latter is the inverse of the optical depth squared
across a scale $\sim k^{-1}$.  With $G=0$, 
equation (\ref{general_flux}) is identical to eq.~[101.62] of 
Mihalas \& Mihalas (1984).\footnote{Correcting for a sign error in 
the first term of their eq.~[101.58].}
For large optical depth in a radiation pressure dominated 
medium, $\nu^2/\Gamma^2\ll1\ll s$ in the third and fifth terms of 
equation (\ref{general_flux}).  
Additionally,  for a non-relativistic 
medium $\nu^2\gg (4 u_r/9\rho) k^2$, $c_T^2k^2$, and $4\pi G\rho$ in the 
fourth term in equation (\ref{general_flux}). 
Using this ordering, dividing through by the quantity $-\Gamma^2s^2$, 
and noting that $\nu^2/\Gamma=ck^2/(3\kappa\rho)$ 
(cf.~eqs.~[\ref{rad_diff}] \& [\ref{gamma}]),
equation (\ref{general_flux}) becomes
\beq
-\left[\frac{\omega^5}{\Gamma^2s^2}\right]-\left[\frac{\omega^4}{i\Gamma s}\right]
+\omega^3
+i\frac{ck^2}{3\kappa\rho}\omega^2
-\omega\left(\frac{4}{3}c_T^2k^2 +\frac{4}{9}\frac{u_r}{\rho}k^2 -4\pi G\rho\right)
-i\frac{ck^2}{3\kappa\rho}\left(c_T^2 k^2-4\pi G\rho\right)\approx0.
\label{general_limit}
\eeq
This expression should be compared with equation (\ref{simpledispersion}).
Note that the last four terms in equation (\ref{general_limit}) are
qualitatively identical to the terms in  equation (\ref{simpledispersion})
in the radiation pressure dominated limit $(A/C_V)\approx1$.
The importance of the first and second terms in equation (\ref{general_limit})
are measured by the characteristic frequency $\Gamma s$
with respect to the wave frequencies the expression admits.  Generically,
for sufficiently large $\Gamma s/\omega$, these terms are sub-dominant.  Thus,
if these limits hold, the qualitative stability properties outlined in 
\S\ref{section:jeans} obtain.

It is simplest to understand the origin of the extra terms in 
equation (\ref{general_flux}) with respect to equation (\ref{simpledispersion})
by taking a step back.  Neglecting the time-dependence of the flux in the last 
expression in equation (\ref{general_transport}),
but leaving the rest of the above analysis unchanged, I find that
\beqa
\omega^4&+&i\Gamma\omega^3\left(1+s+\frac{\nu^2}{\Gamma^2}\right)
-\omega^2\left[\nu^2s+\gamma c_T^2k^2+
\frac{4}{9}\frac{u_r}{\rho}k^2-4\pi G\rho\right] \nonumber \\
&-&i\Gamma\omega\left[\gamma c_T^2k^2\left(\frac{4s}{3\gamma}+1+\frac{\nu^2}{\Gamma^2}\right)
+\frac{4}{9}\frac{u_r}{\rho}k^2\left(3\gamma-3+s\right)
-4\pi G\rho\left(1+s+\frac{\nu^2}{\Gamma^2}\right)\right] \nonumber \\
&+&s\nu^2\left(c_T^2 k^2-4\pi G\rho\right)=0.
\label{general_noflux}
\eeqa
The somewhat peculiar terms multiplying $\gamma c_T^2k^2$ and 
$(4/9)u_rk^2/\rho$ in the fourth term of the dispersion relation (e.g., ``$4s/3\gamma$'')
are made clear by examining the explicit and full expression for the sound speed
of the radiation and the gas, at constant total entropy (cf.~eq.~[\ref{relation_adiabatic}]).
When $u_r/\rho\gg c_T^2$, one finds that $c_s^2\approx(4/3)c_T^2+(4/9)u_r/\rho$,
whereas when  $u_r/\rho\ll c_T^2$, one finds that 
$c_s^2\approx\gamma c_T^2+(4/3)(\gamma-1)u_r/\rho$.  This shows that for large
and small $s$, these terms reduce to the adiabatic sound speed for the radiation
and the gas, respectively.
As in deriving equation (\ref{general_limit}), if I take 
$\nu^2/\Gamma^2\ll1\ll s$ (second and fourth terms) 
and $\nu^2\gg (4 u_r/9\rho) k^2$, $c_T^2k^2$, and $4\pi G\rho$ (third term),
I find that equation (\ref{general_noflux}) can be written simply as
\beq
-\left[\frac{\omega^4}{i\Gamma s}\right]
+\omega^3
+i\frac{ck^2}{3\kappa\rho}\omega^2
-\omega\left(\frac{4}{3}c_T^2k^2 +\frac{4}{9}\frac{u_r}{\rho}k^2 -4\pi G\rho\right)
-i\frac{ck^2}{3\kappa\rho}\left(c_T^2 k^2-4\pi G\rho\right)=0.
\label{general_limit_noflux}
\eeq
Compare with equation (\ref{general_limit}).  The importance of $\Gamma s$ is again evident.
As emphasized by Bogan et al.~(1996) and Blaes \& Socrates (2003), the characteristic frequency 
\beq
\omega_{\rm th}=\Gamma s=\kappa\rho c\left(\frac{4aT^4}{u_g}\right)
\label{omegath}
\eeq
measures the rate at which energy is exchanged between the matter and the 
radiation field.  When this frequency is large, the energetic coupling
is tight and the analysis presented in \S\ref{section:jeans} is recovered.  Thus,
for very large $\Gamma s$ and vanishingly small diffusion rate
across a scale $k^{-1}$ --- $\nu^2/\Gamma=c k^2/(3\kappa\rho)\rightarrow 0$ ---
only the second and third terms in equation (\ref{general_limit_noflux}) survive:
$\omega^3-\omega(4 c_T^2k^2/3+(4/9)u_rk^2/\rho-4\pi G\rho)\approx0$. 
That is, in the optically-thick limit with slow diffusion the 
radiation pressure contributes to the stability of the system against gravitational
collapse.  As in \S\ref{section:jeans}, the fact that it appears
that the system is stabilized if $(4/9)u_rk^2/\rho>4\pi G\rho$,
even when $(4/3)c_T^2k^2<4\pi G\rho$ is an artifact of taking the
limit of zero diffusion rate.
Taking just the last two terms in equation  (\ref{general_noflux})
(the small $\omega$ limit) and then taking $\nu^2/\Gamma^2\rightarrow0$ 
(high optical depth on a scale $k^{-1}$) and
then $s\rightarrow\infty$ (for tight energetic coupling between
the radiation and the matter, radiation pressure-dominated),
I find that 
\beq
\omega\approx\,\,\,-i\frac{\nu^2}{\Gamma}\,\,\displaystyle\left[\displaystyle\frac{c_T^2k^2-4\pi G\rho}
{(4/3)c_T^2k^2+(4/9)(u_r/\rho)k^2-4\pi G\rho}\right]=
i\frac{ck^2}{3\kappa\rho}\,\,\left(\frac{1-\vartheta_T}
{(4/3)\vartheta_T+\vartheta_r-1}\right),
\label{large_gamma_limit}
\eeq
where 
\beq
\vartheta_r=\frac{4}{9}\frac{u_rk^2}{\rho}\frac{1}{(4\pi G\rho)}
\label{vartheta_r}
\eeq
is defined in analogy with $\vartheta_T$
and $\vartheta_s$ (eqs.~[\ref{beta}] \& [\ref{delta}]).  This expression should
be compared with equation (\ref{small_gamma_unstable}); in the radiation pressure
dominated limit they are identical.
Thus, as in \S\ref{section:jeans}, I find that even for highly radiation pressure dominated
media, with very large optical depth and strong energetic coupling between
matter and radiation, the medium is unstable if $c_T^2k^2<4\pi G\rho$ 
(the isothermal Jeans number $\vartheta_T<1$).  As before, at
high-$k$, the characteristic timescale for instability is independent
of spatial scale and is simply the Kelvin-Helmholz timescale
(cf.~eqs.~[\ref{slow_collapse}] \&  [\ref{simple_growth}]).

Alternatively, taking the limit $\Gamma\rightarrow0$ in equation (\ref{general_noflux}), 
so that $\nu^2/\Gamma^2\gg s$, $\nu^2/\Gamma^2\gg1$, $\nu^2/\Gamma^2\gg\nu^2 s$
equation (\ref{general_noflux}) becomes 
\beq
\omega^3-\omega(\gamma c_T^2 k^2-4\pi G\rho)\approx0.
\label{small_gamma_limit}
\eeq
Contrary to the discussion of \S\ref{section:jeans}, which showed
that in the limit of rapid diffusion the gas acoustic mode speed
is the isothermal sound speed $c_T$, here I find that when the 
energetic coupling between the radiation and the gas is weak, 
acoustic modes propagate at the adiabatic sound speed $\gamma^{1/2}c_T$, as expected
(e.g., Mihalas \& Mihalas 1984).
This effect was not accounted for in the analysis of \S\ref{section:jeans}
because perfect energetic coupling was assumed.  Equation (\ref{small_gamma_limit})
shows that in the limit $\Gamma\rightarrow0$, the classical gas
Jeans criterion is obtained and that on scales larger than the Jeans
length, the medium is unstable, even if $(4/9)u_rk^2/\rho\gg4\pi G\rho$.
However, equation (\ref{small_gamma_limit}) is somewhat deceiving as
it may imply to the reader that the medium can be stabilized in the 
special case $c_T^2k^2<4\pi G\rho$, {\it but} $\gamma c_T^2k^2>4\pi G\rho$
in the $\Gamma\rightarrow0$ limit.  This is false, and an artifact 
of having taken $\Gamma=0$ in obtaining equation (\ref{small_gamma_limit}).
Expanding instead to first order in $\Gamma$, I find the unstable mode is
\beq
\omega\approx-i\Gamma s\left(\frac{c_T^2k^2-4\pi G\rho}{\gamma c_T^2k^2-4\pi G\rho}\right)
=i\Gamma s\left(\frac{1-\vartheta_T}{\gamma\vartheta_T-1}\right),
\label{unstable_small_gamma}
\eeq
which shows that in the special case $c_T^2k^2<4\pi G\rho$, but
$\gamma c_T^2k^2>4\pi G\rho$, the medium is unstable.

Expanding equation (\ref{general_noflux}) in the high-$k$ limit I find that 
\beq
\omega\approx\pm \gamma^{1/2}c_Tk
-i\frac{3\Gamma}{2}\left[\left(\frac{4}{9}\frac{u_r}{\rho c^2}\right)+
\frac{s}{3}\left(\frac{\gamma-1}{\gamma}\right)\right],
\label{blaes_socrates_general}
\eeq
in agreement with Blaes \& Socrates (2003) (their eq.~[57]).  In a 
non-relativistic medium, the second term in square brackets dominates
so that $\omega\approx\pm\gamma^{1/2}c_Tk-i\Gamma s(\gamma-1)/2\gamma$.
Thus, in the high-$k$ limit gas acoustic waves are damped by emission and absorption,
again with characteristic damping rate $\sim \Gamma s$.
Although in equation (\ref{blaes_socrates_general}) I
obtain a wave speed equal to the 
adiabatic gas sound speed, as in equation (\ref{small_gamma_limit}),
the high-$k$ limit is not identical to the limit $\Gamma\rightarrow0$ because 
the gravitational term, which dictates stability/instability, disappears 
at high $k$.  To see this, I write 
$\omega=\pm\sqrt{\gamma c_T^2k^2-4\pi G\rho}+iq$ in equation (\ref{general_noflux}),
take only linear terms in $q$, and then expand to first order as $\Gamma\rightarrow0$.
I find that  
\beq
\omega\approx\pm (\gamma c_T^2 k^2-4\pi G\rho)^{1/2}
-i\frac{3\Gamma}{2}\left[\left(\frac{4}{9}\frac{u_r}{\rho c^2}\right)+
\frac{s}{3}\frac{(\gamma-1)c_T^2k^2}{\gamma c_T^2k^2-4\pi G\rho}\right].
\label{grav_noflux_small_gamma}
\eeq
For large $k$, equation (\ref{grav_noflux_small_gamma}) reduces to 
equation (\ref{blaes_socrates_general}).  However, on scales
where gravity is important, equation (\ref{grav_noflux_small_gamma})
shows that if $\gamma c_T^2k^2\gtrsim4\pi G\rho$ then the damping 
rate of gravity-modified adiabatic gas acoustic waves is altered
from the prediction of equation (\ref{blaes_socrates_general}).
More importantly, we see explicitly that if $\gamma c_T^2k^2<4\pi G\rho$
the acoustic mode is unstable.  

Taken together, equations (\ref{unstable_small_gamma}) and (\ref{grav_noflux_small_gamma})
show that in the limit of small $\Gamma s$, 
(1) if {\it both} $\vartheta_T<1$ and $\gamma\vartheta_T<1$, then 
the medium is unstable, and (2) that if $\vartheta_T<1$ and $\gamma\vartheta_T>1$, 
then the medium is {\it also} unstable.   In addition, equation (\ref{large_gamma_limit})
shows that when $\Gamma s$ is very large and the energetic coupling between
matter and radiation is tight, the medium is {\it also} unstable
for $\vartheta_T<1$, even if $\vartheta_r\gg1$, as in \S\ref{section:jeans}.  
Thus, one expects that 
the medium is only globally stable on a scale $k^{-1}$ if the
classical Jeans criterion is satisfied and the isothermal Jeans number 
is larger than unity, $\vartheta_T>1$.  

\subsection{Full Solution to the Dispersion Relation}

In analogy with $\vartheta_T$, $\vartheta_s$, $\chi$, and $\vartheta_r$ 
(eqs.~[\ref{beta}]$-$[\ref{gamma}] \& [\ref{vartheta_r}]), I define the quantities
\beq
\alpha=\frac{c^2k^2}{3}\frac{1}{(4\pi G\rho)}=\frac{\nu^2}{(4\pi G\rho)},
\hspace{.5cm}
{\rm and} \,\,\,\,\,\zeta=\frac{\Gamma s}{(4\pi G\rho)^{1/2}},
\eeq
where $\zeta$ measures the rate of thermal coupling between radiation
and gas in units of the dynamical timescale.  With these definitions,
the approximate solutions and limits of the previous sub-section
are illustrated in Figure \ref{fig:gen_nof},
which presents the full solution to equation (\ref{general_noflux})
over a broad range of $\nu/\Gamma\propto\tau_k^{-1}$ for 
the parameters $\alpha=10^7$, $\gamma=5/3$, 
$\vartheta_r=3$, 
$\vartheta_T=1/5$ ({\it top left panel}), 
$\vartheta_T=1/2$ ({\it top right panel}),
$\vartheta_T=4/5$ ({\it bottom left panel}), and
$\vartheta_T=3/2$ ({\it bottom right panel}) at fixed scale $k^{-1}$.
In each panel, I take $s=9(\gamma-1)\vartheta_r/\vartheta_T$
so that $s=180$, 36, 22.5, and 12 from left to right, top to bottom,
respectively.
As in Figures \ref{fig:simple} and \ref{fig:simplel},
open and filled circles show the real and imaginary components
of $\xi=\omega/(4\pi G\rho)^{1/2}$.  Positive complex components
indicate unstable modes.   Note that in each panel the damping rates
$\xi\approx -i\chi$ and $\xi\approx -i\zeta$ are off-scale at intermediate
values of $\nu/\Gamma$.

In each panel, because $\vartheta_r$ --- and, by extension, $\vartheta_s$ ---
is larger than unity, in the limit of large $\tau_k$  (large $\Gamma$, small $\nu/\Gamma$)
solutions qualitatively identical to those obtained in 
Figures \ref{fig:simple} and \ref{fig:simplel} are recovered.  Thus, the left-hand
portions of all panels are similar to Figure \ref{fig:simple}.  The 
qualitatively new feature of this figure is the small-$\tau_k$ regions
in each panel.
In the top two panels both $c_T^2k^2$ and $\gamma c_T^2 k^2$ are less than $4\pi G\rho$
($\vartheta_T<1$ and $\gamma \vartheta_T<1$) so that the medium is unstable 
for any $\nu/\Gamma$.
For sufficiently small $\nu/\Gamma$, the 
growth rate is sub-dynamical and equal to the inverse of the 
Kelvin-Helmholz timescale (cf.~eqs.~[\ref{slow_collapse}] \& [\ref{simple_growth}]). 
Here, the diffusive instability operates.
However, for intermediate values of the $\nu/\Gamma$,
the growth rate is dynamical ($\xi\approx\pm i(1-\vartheta_T)^{1/2}$), whereas
for very large $\nu/\Gamma$, $\xi\approx\pm i(1-\gamma\vartheta_T)^{1/2}$. 
In these cases, the medium is classically Jeans unstable.

The bottom left panel is different.  Here, 
$\vartheta_T<1$, but $\gamma \vartheta_T>1$ (see eq.~[\ref{unstable_small_gamma}]).  
In this special case, two sets of gravity- and diffusion- modified acoustic waves
exist: (1) at small $\nu/\Gamma$, the adiabatic radiation pressure dominated acoustic 
waves are evident (as in the top two panels) and (2) at large $\nu/\Gamma$, 
adiabatic gas (only) acoustic waves are also present.  
Note that in this regime  (small thermal coupling,
small $\tau_k$) the medium is still unstable, but the growth rate for
instability is 
$\approx\zeta(1-\vartheta_T)(\gamma\vartheta_T-1)^{-1}$ (eq.~[\ref{unstable_small_gamma}])
in the special case
where both the numerator and the denominator are positive.

The bottom right panel shows a case analogous to the right panels of 
Figures \ref{fig:simple} and \ref{fig:simplel} with $\vartheta_T>1$.
Here, the Jeans instability is stabilized at all $\tau_k$.  At intermediate
$\nu/\Gamma$, $\xi\approx\pm(\vartheta_T-1)^{1/2}$, whereas for 
large values of $\nu/\Gamma$,  $\xi\approx\pm(\gamma\vartheta_T-1)^{1/2}$.
As in the other panels, in the limit of strong thermal coupling between
the radiation and the gas (large $\tau_k$, small $\nu/\Gamma$),
$\xi\approx\pm(\vartheta_s-1)^{1/2}$.

\subsection{Scalings \& Applications}

\begin{figure}
\centerline{\includegraphics[width=9cm]{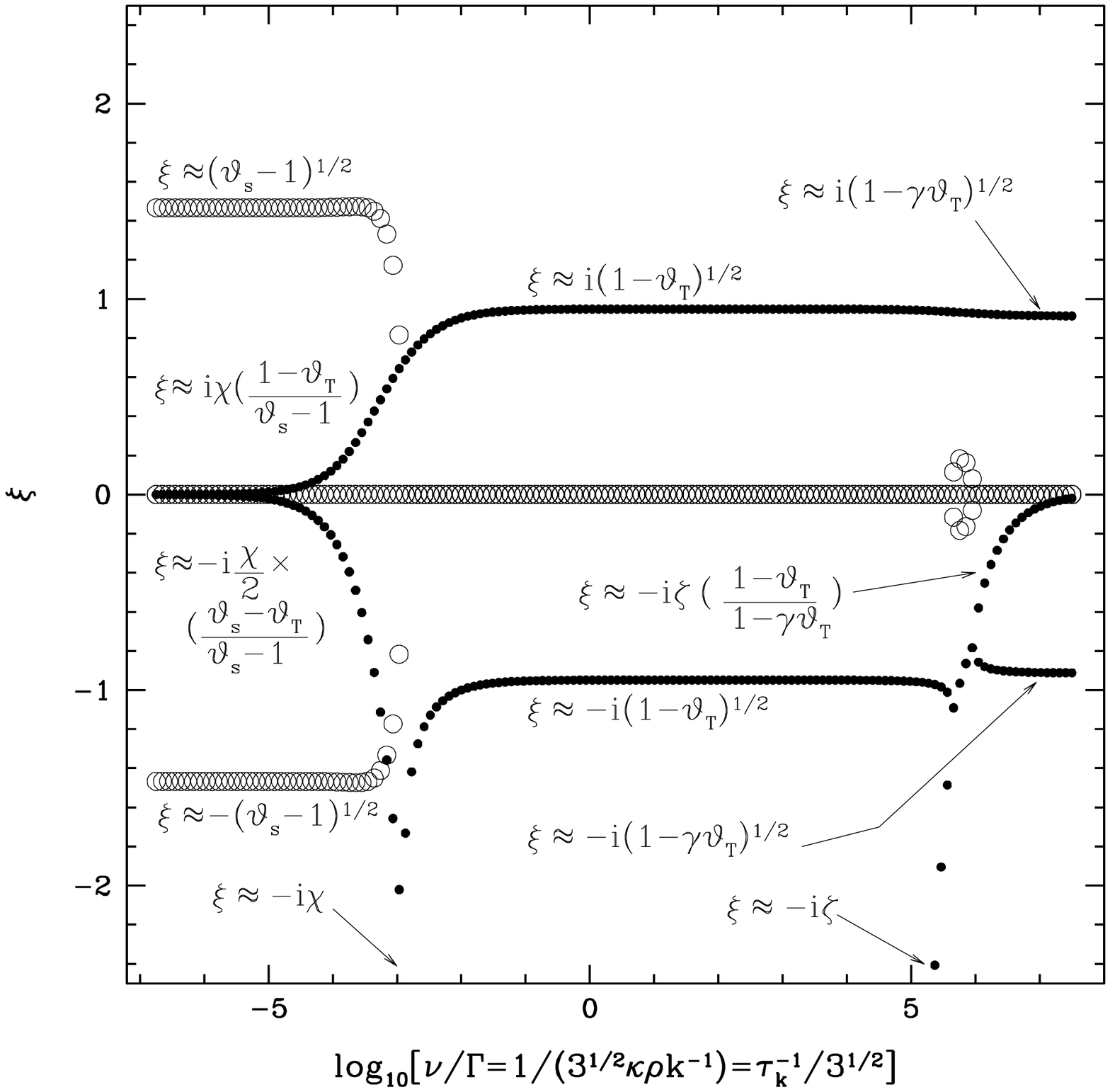}\includegraphics[width=9cm]{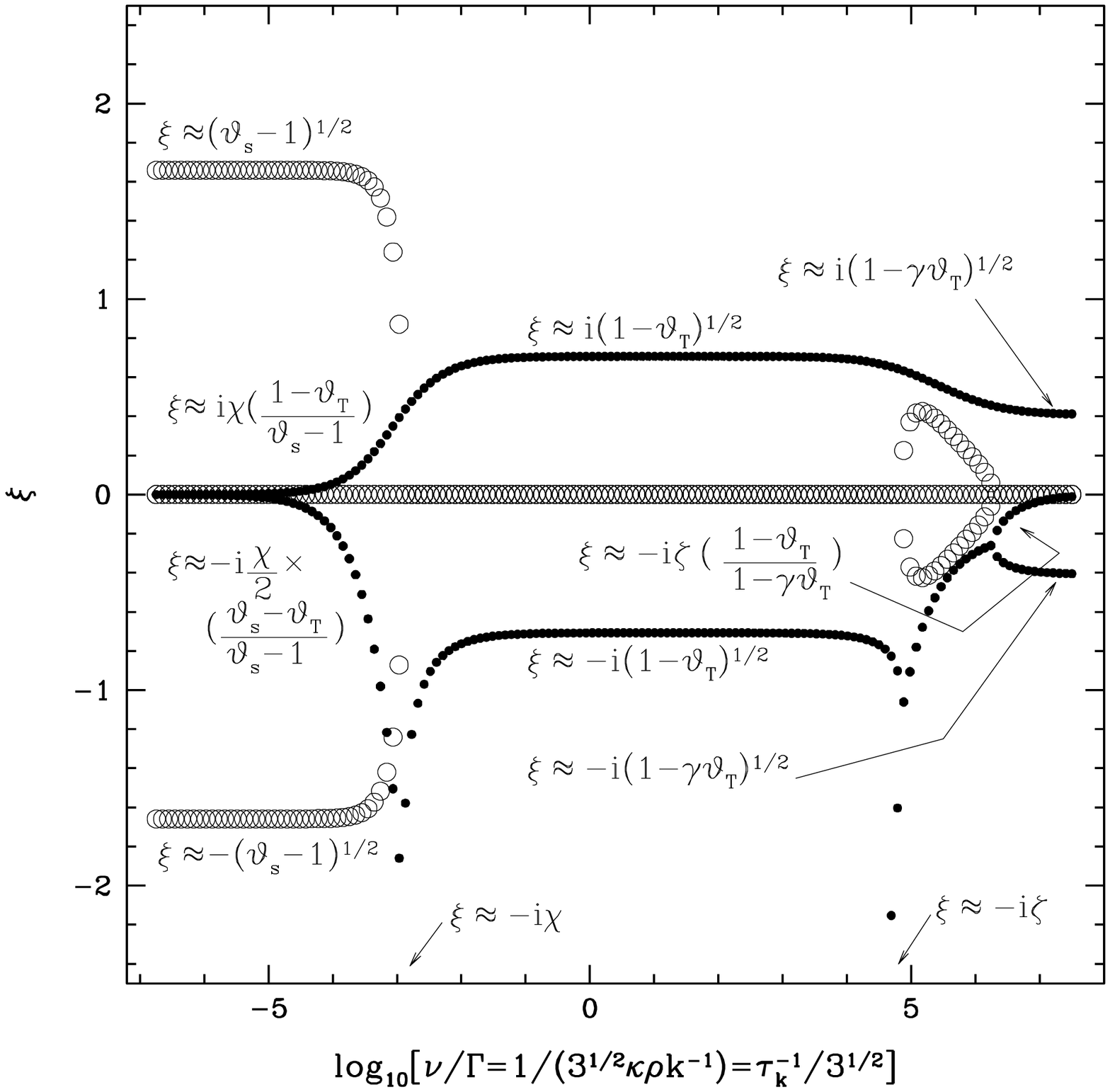}}
\centerline{\includegraphics[width=9cm]{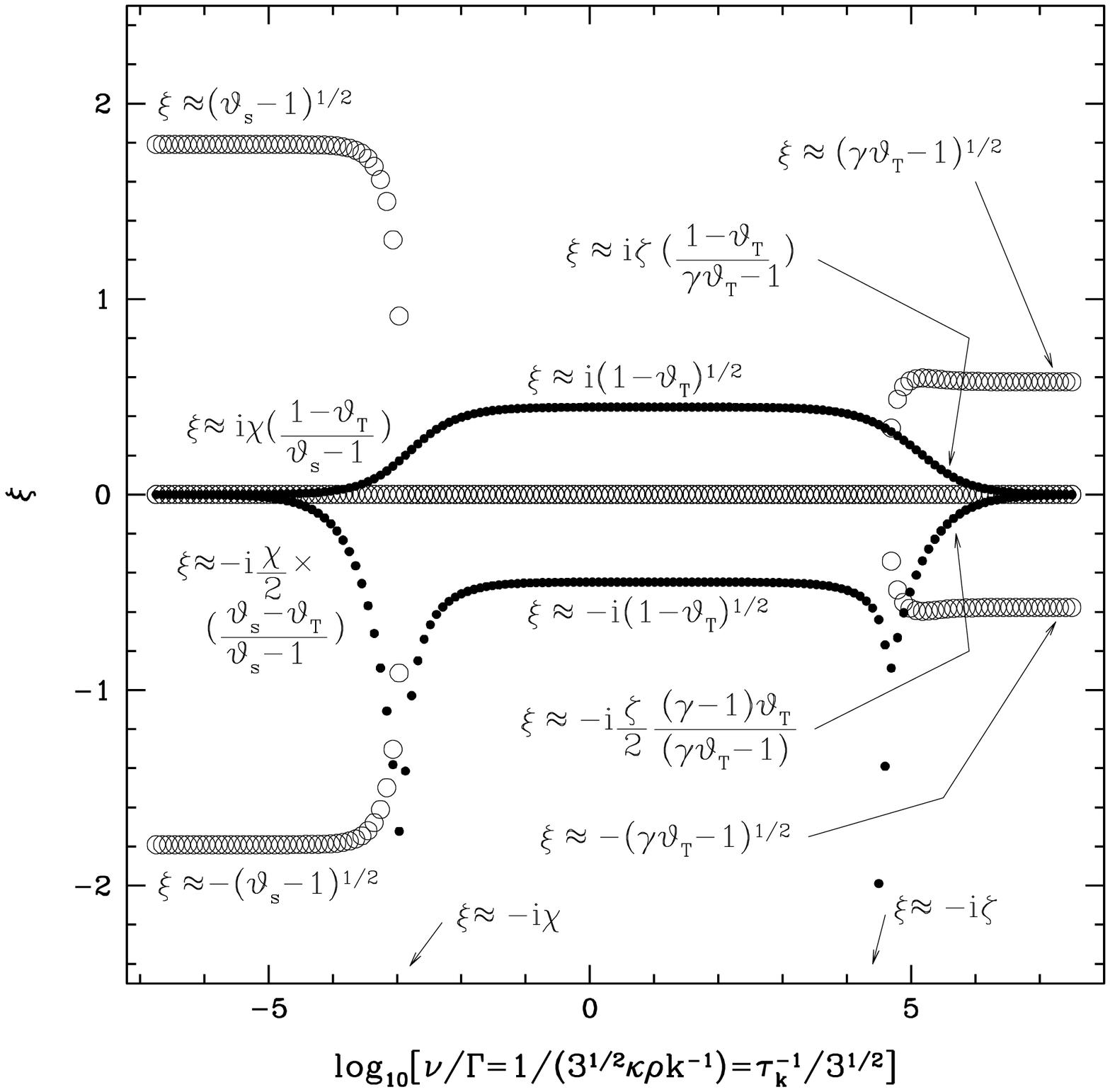}\includegraphics[width=9cm]{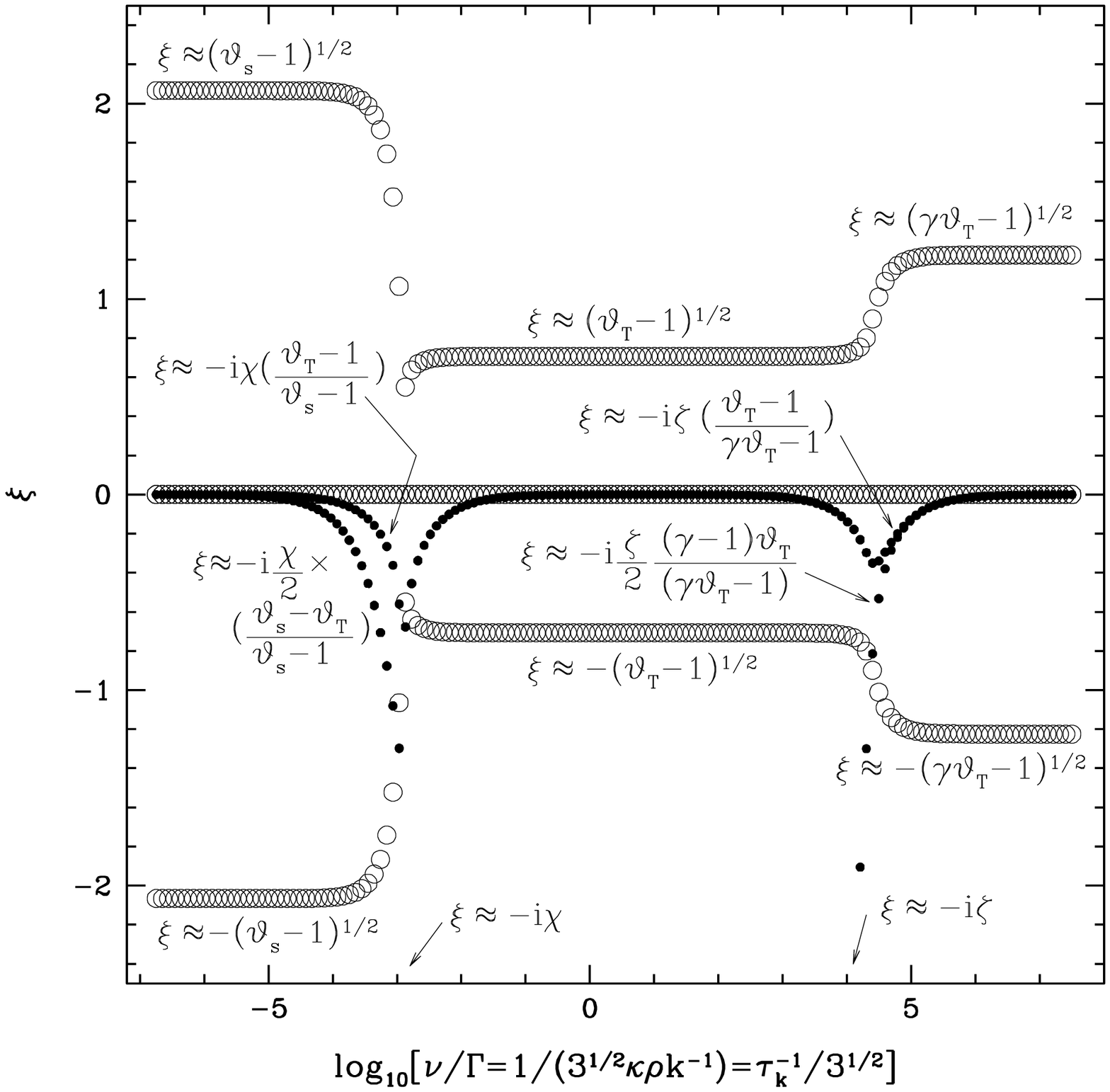}}
\figcaption[general_nof]{Solution to equation (\ref{general_noflux})
for $\alpha=10^7$, $\gamma=5/3$, $\vartheta_r=3$, 
$\vartheta_T=1/10$ ({\it top left panel}) and for $\vartheta_T=1/2$ ({\it top right panel}),
$\vartheta_T=4/5$ ({\it bottom left panel}), $\vartheta_T=3/2$ ({\it bottom right panel}),
for a very wide range of $\nu/\Gamma=\tau_k^{-1}/3^{1/2}$ at fixed scale $k^{-1}$.
Open and filled circles show the real and imaginary parts of the roots $\xi$, 
respectively. Positive complex roots indicate instability.  Components of the 
dispersion relation are labeled for clarity. The left portion of
each panel (large $\tau_k$, strong thermal coupling) is qualitatively similar to 
Figures \ref{fig:simple} and \ref{fig:simplel}.  In each panel $\vartheta_r>1$
so that $\vartheta_s>1$.  The bottom left panel shows the special case
$\vartheta_T<1$ and $\gamma\vartheta_T>1$.  At very large $\nu/\Gamma$, the 
gravity- and radiation-modified adiabatic {\it gas only} acoustic wave 
exists and in this region the growth rate for the Jeans instability is suppressed.
\label{fig:gen_nof}}
\end{figure}

The qualitative differences at small optical depth or
poor thermal coupling between the radiation and the matter in the 
right hand portion of each of the panels in Figure \ref{fig:gen_nof} 
with respect to Figures \ref{fig:simple} and \ref{fig:simplel} 
do not change the conclusions about most of the astrophysical
applications discussed in \S\ref{section:discussion}
because fiducial estimates for $\zeta$ and $\alpha$ are very 
large:
%
\beqa
\zeta=
\frac{\Gamma s}{(4\pi G\rho)^{1/2}}
&=&\frac{\kappa\rho c}{(4\pi G\rho)^{1/2}}\left(\frac{4aT^4}{u_g}\right)
\,\,\approx\,\,10^{8}\kappa_{2.5}n_4^{-1/2}T_{2}^3
\,\,\approx\,\,10^{9}\kappa_{2.5}n_8^{-1/2}T_{3}^3 \nonumber \\
&=&\frac{c}{l_{\rm mfp}}\frac{12(\gamma-1)}{(4\pi G\rho)^{1/2}}\left(\frac{p_{\rm cr}}{p_g}\right)
\approx\,\,7\times10^{8}l_{\rm 0.1\,pc}^{-1}n^{-1/2}\left(\frac{p_{\rm cr}}{p_g}\right)
\label{scale_omegath}
\eeqa
and
\beqa
\alpha=\frac{\nu^2}{4\pi G\rho}=
\frac{c^2k^2/3}{4\pi G\rho}\approx\frac{c^2\pi}{G\rho\lambda^2}
\,\,&\approx&\,\,7\times10^6\lambda_{2}^{-2}n_4^{-1}
\,\,\approx\,\,7\times10^8\lambda_{-1}^{-2}n_8^{-1} \nonumber \\
&\approx&\,\,7\times10^8\lambda_{\rm kpc}^{-2}n^{-1}.
\eeqa
Additionally, the ratio $\nu/\Gamma$ is 
\beqa
\frac{\nu}{\Gamma}=\frac{1}{3^{1/2}}\left(\frac{k}{\kappa\rho}\right)
&=&\frac{1}{3^{1/2}}\left(\frac{2\pi}{\kappa\rho\lambda}\right)
\,\,\approx\,\,0.3\kappa_{2.5}n_4^{-1}\lambda_{2}^{-1}
\,\,\approx\,\,0.03\kappa_{2.5}n_8^{-1}\lambda_{-1}^{-1} \nonumber \\
&=&\frac{1}{3^{1/2}}\left(\frac{2\pi l_{\rm mfp}}{\lambda}\right)
\,\,\approx\,\,4\times10^{-4}l_{0.1\,{\rm pc}}\lambda_{\rm kpc}^{-1}.
\eeqa
The first line of each equation shows the scalings for starbursts and AGN disks,
while the second line shows the scaling appropriate for cosmic rays
(cf.~eq.~[\ref{dimen_cr}]).  For all cases
considered, $\zeta$ is very large and $\nu/\Gamma$ is less than unity.  
The stability properties of these media are thus best represented by the 
left-hand portion of each of the panels in Figure \ref{fig:gen_nof};
the radiation and the matter are tightly energetically coupled.

\clearpage
\clearpage




\begin{thebibliography}{}

\bibitem[Agol \& Krolik(1998)]{1998ApJ...507..304A} 
Agol, E., \& Krolik, J.\ 1998, \apj, 507, 304 

\bibitem[Arons(1992)]{1992ApJ...388..561A} Arons, J.\ 1992, \apj, 388, 561 

\bibitem[Binney \& Tremaine(1987)]{1987gady.book.....B} Binney, J., \& 
Tremaine, S.\ 1987, Princeton, NJ, Princeton University Press, 1987

\bibitem[Bisnovatyi-Kogan \& Blinnikov(1979)]{1979Ap.....15...99B} 
Bisnovatyi-Kogan, G.~S., \& Blinnikov, S.~I.\ 1979, Astrophysics, 15, 99 

\bibitem[Bisnovatyi-Kogan \& Blinnikov(1978)]{1978Ap.....14..316B} 
Bisnovatyi-Kogan, G.~S., \& Blinnikov, S.~I.\ 1978, Astrophysics, 14, 316 

\bibitem[Blaes \& Socrates(2001)]{2001ApJ...553..987B} 
Blaes, O., \& Socrates, A.\ 2001, \apj, 553, 987 

\bibitem[Blaes \& Socrates(2003)]{2003ApJ...596..509B} 
Blaes, O., \& Socrates, A.\ 2003, \apj, 596, 509 

\bibitem[Bogdan et al.(1996)]{1996ApJ...456..879B} Bogdan, T.~J., Knoelker, 
M., MacGregor, K.~B., \& Kim, E.-J.\ 1996, \apj, 456, 879 

\bibitem[Boulares \& Cox(1990)]{1990ApJ...365..544B} 
Boulares, A., \& Cox, D.~P.\ 1990, \apj, 365, 544 

\bibitem[Chandrasekhar \& Fermi(1953)]{1953ApJ...118..116C} 
Chandrasekhar, S., \& Fermi, E.\ 1953, \apj, 118, 116 

\bibitem[Chandrasekhar(1954)]{1954ApJ...119....7C} 
Chandrasekhar, S.\ 1954, \apj, 119, 7 

\bibitem[Chandrasekhar(1961)]{1961hhs..book.....C} Chandrasekhar, S.\ 1961, 
International Series of Monographs on Physics, Oxford: Clarendon, 1961

\bibitem[Chang et al.(2006)]{2006astro.ph.10263C} Chang, P., Quataert, E., 
\& Murray, N.\ 2006, ArXiv Astrophysics e-prints, arXiv:astro-ph/0610263 

\bibitem[Condon et al.(1991)]{1991ApJ...378...65C} Condon, J.~J., Huang, 
Z.-P., Yin, Q.~F., \& Thuan, T.~X.\ 1991, \apj, 378, 65 

\bibitem[Condon(1992)]{1992ARA&A..30..575C} Condon, J.~J.\ 1992, \araa, 30, 575 

\bibitem[Connell(1998)]{1998ApJ...501L..59C} 
Connell, J.~J.\ 1998, \apjl, 501, L59 

\bibitem[Dodelson(2003)]{2003moco.book.....D} Dodelson, S.\ 2003, Modern 
cosmology / Scott Dodelson.~Amsterdam (Netherlands): Academic Press., 2003

\bibitem[Downes \& Solomon(1998)]{1998ApJ...507..615D} Downes, D., \& 
Solomon, P.~M.\ 1998, \apj, 507, 615 

\bibitem[Dzhalilov et al.(1992)]{1992A&A...257..359D} Dzhalilov, N.~S., 
Zhugzhda, Y.~D., \& Staude, J.\ 1992, \aap, 257, 359 

\bibitem[Gammie(1998)]{1998MNRAS.297..929G} 
Gammie, C.~F.\ 1998, \mnras, 297, 929 

\bibitem[Garcia-Munoz et al.(1977)]{1977ApJ...217..859G} 
Garcia-Munoz, M., Mason, G.~M., \& Simpson, J.~A.\ 1977, \apj, 217, 859 

\bibitem[Goldreich \& Lynden-Bell(1965)]{1965MNRAS.130...97G} 
Goldreich, P., \& Lynden-Bell, D.\ 1965, \mnras, 130, 97 

\bibitem[Gorti \& Hollenbach(2004)]{2004ApJ...613..424G} Gorti, U., \& 
Hollenbach, D.\ 2004, \apj, 613, 424 

\bibitem[Goodman(2003)]{2003MNRAS.339..937G} 
Goodman, J.\ 2003, \mnras, 339, 937 

\bibitem[Guo \& Oh(2008)]{2008MNRAS.384..251G} Guo, F., \& Oh, S.~P.\ 2008, \mnras, 384, 251 

\bibitem[Hansen(1978)]{1978ARA&A..16...15H} Hansen, C.~J.\ 1978, \araa, 16, 15 

\bibitem[Hu \& Sugiyama(1996)]{1996ApJ...471..542H} Hu, W., \& Sugiyama, 
N.\ 1996, \apj, 471, 542 

\bibitem[Iwasawa et al.(2001)]{2001MNRAS.326..894I} Iwasawa, K., Matt, G., 
Guainazzi, M., \& Fabian, A.~C.\ 2001, \mnras, 326, 894 

\bibitem[Jeans(1928)]{1928QB43.J4........} 
Jeans, J.~H.\ 1928, Cambridge [Eng.] The University press, 1928.

\bibitem[Kaneko et al.(1976)]{1976Ap&SS..42..441K} Kaneko, N., Tamazawa, 
S., \& Ono, Y.\ 1976, \apss, 42, 441 

\bibitem[Kaneko \& Morita(2006)]{2006Ap&SS.305..349K} Kaneko, N., \& 
Morita, K.\ 2006, \apss, 305, 349 

\bibitem[Kennicutt(1998)]{1998ApJ...498..541K} 
Kennicutt, R.~C., Jr.\ 1998, \apj, 498, 541 

\bibitem[Klessen et al.(2000)]{2000ApJ...535..887K} Klessen, R.~S., 
Heitsch, F., \& Mac Low, M.-M.\ 2000, \apj, 535, 887 

\bibitem[Krolik(2007)]{2007ApJ...661...52K} Krolik, J.~H.\ 2007, \apj, 661, 
52 

\bibitem[Krumholz \& McKee(2005)]{2005ApJ...630..250K} 
Krumholz, M.~R., \& McKee, C.~F.\ 2005, \apj, 630, 250 

\bibitem[Kuwabara \& Ko(2004)]{2004JKAS...37..601K} Kuwabara, T., \& Ko, 
C.-M.\ 2004, Journal of Korean Astronomical Society, 37, 601 

\bibitem[Ledoux(1951)]{1951AnAp...14..438L} 
Ledoux, P.\ 1951, Annales d'Astrophysique, 14, 438 

\bibitem[Lithwick \& Goldreich(2001)]{2001ApJ...562..279L} Lithwick, Y., \& 
Goldreich, P.\ 2001, \apj, 562, 279 

\bibitem[Lynden-Bell(1966)]{1966Obs....86...57L} 
Lynden-Bell, D.\ 1966, The Observatory, 86, 57 

\bibitem[Mac Low \& Klessen(2004)]{2004RvMP...76..125M} Mac Low, M.-M., \& 
Klessen, R.~S.\ 2004, Reviews of Modern Physics, 76, 125 

\bibitem[Mihalas \& Mihalas(1983)]{1983ApJ...273..355M} Mihalas, D., \& 
Mihalas, B.~W.\ 1983, \apj, 273, 355 

\bibitem[Mihalas \& Mihalas(1984)]{book} Mihalas, D., \& 
Mihalas, B.~W.\ 1984, New York, NY, Oxford University Press, 1984


\bibitem[Mestel(1965)]{1965QJRAS...6..161M} Mestel, L.\ 1965, \qjras, 6, 161 

\bibitem[Peebles \& Yu(1970)]{1970ApJ...162..815P} 
Peebles, P.~J.~E., \& Yu, J.~T.\ 1970, \apj, 162, 815 

\bibitem[Pier \& Krolik(1992)]{1992ApJ...399L..23P} Pier, E.~A., \& Krolik, 
J.~H.\ 1992, \apjl, 399, L23 

\bibitem[Scoville(2003)]{2003JKAS...36..167S} Scoville, N.\ 2003, Journal 
of Korean Astronomical Society, 36, 167 

\bibitem[Scoville et al.(2001)]{2001AJ....122.3017S} Scoville, N.~Z., 
Polletta, M., Ewald, S., Stolovy, S.~R., Thompson, R., \& Rieke, M.\ 2001, 
\aj, 122, 3017 

\bibitem[Semenov et al.(2003)]{2003A&A...410..611S} Semenov, D., Henning, 
T., Helling, C., Ilgner, M., \& Sedlmayr, E.\ 2003, \aap, 410, 611 

\bibitem[Silk(1967)]{1967Natur.215.1155S} 
Silk, J.\ 1967, \nat, 215, 1155 

\bibitem[Silk(1968)]{1968ApJ...151..459S} Silk, J.\ 1968, \apj, 151, 459 

\bibitem[Sirko \& Goodman(2003)]{2003MNRAS.341..501S} 
Sirko, E., \& Goodman, J.\ 2003, \mnras, 341, 501 


\bibitem[Socrates et al.(2006)]{2006astro.ph..9796S} Socrates, A., Davis, 
S.~W., \& Ramirez-Ruiz, E.\ 2006, arXiv:astro-ph/0609796 

\bibitem[Spiegel(1957)]{1957ApJ...126..202S} Spiegel, E.~A.\ 1957, \apj, 
126, 202 

\bibitem[Thompson et al.(2005)]{2005ApJ...630..167T} 
Thompson, T.~A., Quataert, E., \& Murray, N.\ 2005, \apj, 630, 167 [TQM]

\bibitem[Thompson et al.(2006)]{2006ApJ...645..186T} Thompson, T.~A., 
Quataert, E., Waxman, E., Murray, N., \& Martin, C.~L.\ 2006, \apj, 645, 
186

\bibitem[Toomre(1964)]{1964ApJ...139.1217T} 
Toomre, A.\ 1964, \apj, 139, 1217 

\bibitem[Turner et al.(2000)]{2000ApJ...532L.109T} Turner, J.~L., Beck, 
S.~C., \& Ho, P.~T.~P.\ 2000, \apjl, 532, L109 

\bibitem[Turner et al.(2005)]{2005ApJ...624..267T} Turner, N.~J., Blaes, 
O.~M., Socrates, A., Begelman, M.~C., \& Davis, S.~W.\ 2005, \apj, 624, 267 

\bibitem[Vranjes(1990)]{1990Ap&SS.173..293V} 
Vranjes, J.\ 1990, \apss, 173, 293
 
\bibitem[Vranjes \& Cadez(1990)]{1990Ap&SS.164..329V} 
Vranjes, J., \& Cadez, V.\ 1990, \apss, 164, 329 
 
\bibitem[Weinberg(1971)]{1971ApJ...168..175W} 
Weinberg, S.\ 1971, \apj, 168, 175 

\bibitem[Zhugzhda et al.(1993)]{1993A&A...278L...9Z} Zhugzhda, Y.~D., 
Dzhalilov, N.~S., \& Staude, J.\ 1993, \aap, 278, L9 

\end{thebibliography}
\end{document}